\documentclass[twocolumn]{openjournal}

\usepackage[T1]{fontenc}
\usepackage{ae,aecompl}
\usepackage{natbib}
\usepackage{wasysym}

\usepackage{xcolor}

\begin{document}
\title{Widespread disruption of resonant chains during protoplanetary disk dispersal}
\author{Brad M. S. Hansen}
\email[email]{ hansen@astro.ucla.edu}
\affiliation{ Mani L. Bhaumik Institute for Theoretical Physics,  Department of Physics \& Astronomy, University of California Los Angeles, \\ Los Angeles, CA 90095}
\author{Tze-Yeung Yu }
\affiliation{Mani L. Bhaumik Institute for Theoretical Physics,  Department of Physics \& Astronomy, University of California Los Angeles, \\ Los Angeles, CA 9009
}
\author{Yasuhiro Hasegawa}
\affiliation{Jet Propulsion Laboratory, California Institute of Technology, Pasadena, CA, 91109}

\begin{abstract}
We describe the evolution of low mass planets in a dispersing protoplanetary disk around a Solar mass star. The disk model is based on the results of \cite{YHH23}, which describes a region of the inner disk where the direction of the migration torque is outwards due to the diffusion of the stellar magnetic field into the disk and the resultant gradual increase in surface density outwards. We demonstrate that the magnetospheric  rebound phase in such a disk leads to diverging orbits for double and triple planet systems, and the disruption of a high fraction of the initial resonant chains.

We present simulations of three planet systems, with masses based on the observed triple planet systems observed by the Kepler satellite,
within the context of this model. The final distribution of nearest neighbour period ratios provides an excellent fit to the observations, provided that the initial  configurations contain a significant fraction of pairs whose period ratios are less than  2:1. The occurrence rate of planets as a function of orbital period also provides a good match to the observations, for final orbital periods $P<20$~days.

These results suggest that the period and period ratio distributions of low mass planets are primarily set in place during the disk dispersal epoch, and may not require significant dynamical evolution thereafter.

\end{abstract}

\maketitle


\section{Introduction}

The Kepler satellite mission \citep{Borucki10} has revealed that $>50\%$ of Sun-like stars possess planetary systems containing at least one and, frequently, several planets with orbital periods $< 100$~days \citep{Bat13,Cough16,Liss23}. The majority of these planets have masses that lie between
those of Earth and Saturn, suggesting a commonality with either rocky planets (hereafter Super-Earths) or planets with a substantial
gaseous or water content by mass (hereafter Sub-Neptunes).  Despite their widespread occurrence, the origins of these planets is still
a matter of lively debate.

One school of thought suggests that such planets, especially those with volatile-rich envelopes, formed further from the star and migrated inwards due to interactions with the gaseous protoplanetary disk. The migration timescales associated with low mass planets are quite short, but gravitational interactions between multiple migrating planets can slow the evolution sufficiently to avoid losing the planets into the central star \citep{TP07,IL08,IL10,HN12,Coss14,CN14,Izi17}. The high frequency of multiple planet systems also aligns nicely with this model. However, such models predict a high occurrence rate of mean motion resonances in such systems, which is not matched by observations \citep{Liss11,Fab14}. An alternative class of models suggests that the planets assembled in situ, with little radial migration \citep{HM12,CL13,BMF14,CT15}. These models can better match the observed spacing of planets in multiple systems \citep{HM13}, but require that the protoplanetary disks be either unusually massive or allow for significant inward migration of solid material as small bodies, prior to assembly.

For models in which the planets migrate within the protoplanetary disk, a critical element is the process that determines the halting of migration. In particular, the structure of the inner edge of the disk is expected to be regulated by the interaction with the stellar magnetic field \citep{RL06,RLK19} and the change in the density structure reduces or reverses the torque driving the planets inwards \citep{MMC06,Ts11,ML18}. For the purposes of modelling migration, this 
 is often presented as a simple truncation in surface density, resulting from the balance of the ram pressure of the disk flow with the magnetic pressure of the magnetic field \citep{TP07,CGP10,Izi17,BMI18}. However, \cite{YHH23} have shown that a more accurate model, that accounts for the diffusion of the magnetic field into the disk,  will result in a surface density that increases outwards from the inner edge before reaching a maximum and reverting to a more traditional power-law decay to larger radii.  The change in the surface density and temperature profiles reverses the torque direction interior to the surface density maximum.
 Thus, migrating planets reverse the course of their migration as they cross this transition and  pile up, not at the edge of
the magnetosperic cavity, but at a more distant location -- the surface density maximum. As the accretion rate drops, the torque reversal location moves outwards, causing
the planet to move outwards again, 
 coming to rest at the point where the disk density has dropped to the point that the migration `freezes out'. 

In \cite{YHH23} we demonstrated how this model determines the final orbital periods in the case of a single planet. In this paper, we wish to investigate the consequences of this model for multiple planet systems. An important feature of the multiple planet case is that it can result in planetary pairs breaking out of resonance. \cite{LOL17} have shown that, as the accretion rate drops, the magnetospheric cavity expands and can drive outward migration of a planet trapped at the inner edge. They use a one-sided torque  operating at the disk edge, based on the results of the studies above. The fact that outward torques operate only on one planet in that model limits the degree to which this effect operates to disrupt a resonance chain, and the resulting simulations still leave too many systems in resonant
configurations \citep{LO17}.  As we will show, the presence of a non-zero torque interior to the location of the torque reversal results in multiple planets migrating outwards and provides a more extensive dynamical perturbation to the system.

In this paper we present a calculation of the resonant dynamics of two and three planet systems, evolving under the influence of a torque model adapted from \cite{YHH23}. The dynamical model, including both resonant interactions and torques from the protoplanetary disk, is described in \S\ref{DynModel}. In \S\ref{Times} we adapt the semi-analytic treatment of the resonant dynamics to take into account the torque reversals. \S\ref{Dos} presents the application of our model to two planet systems and describes the character of the evolution over the range of planetary masses. In \S\ref{Tres} we show how the introduction of a third planet influences the dynamics and present a comparison between model results and the Kepler observational sample. The implications of our results are discussed in \S\ref{Disc}.

\section{Dynamical Model}
\label{DynModel}

We wish to model the resonant dynamics of two and three planet systems, under the influence of the most important
resonant interactions. We choose to focus on the first order resonances $(q+1)/q$ where $q=1$ and $q=2$. We will
also include the effects of torques from the protoplanetary disk on both semi-major axis evolution and planetary eccentricity damping.
This is done by including, in the equations, 
a linear decay term characterised by a timescale, as has been done by several prior authors -- \citep{TP19,CC20}. One difference
in our case is that our  torque is based on the model of \cite{YHH23}, which features a reversal of the torque direction in
the inner disk, due to the penetration of the stellar magnetic field into the disk.  
  We must therefore allow for the timescale for semi-major axis evolution ($\tau_a$)
to become negative, in order to allow for outward migration. The timescale for eccentricity damping is always positive. The equations featuring three planets and two
resonances are sufficiently lengthy that we defer  their explicit expression to Appendix~\ref{Eqns}.


\subsection{Torque Model}

The dynamical evolution of the system is driven by the torques exerted by the protoplanetary disk. We determine
a functional form for the torque based on a fit to 
 the numerical disk simulations discussed in \cite{YHH23}. For the inward migration, the characteristic
timescale for semi-major axis evolution of planet $i$ is
\begin{eqnarray}
\tau_{a,i} &=& 0.36 {\rm Myr} \left( \frac{a_i}{0.1 AU}\right)^{1.38} \left( \frac{M_i}{10 M_{\oplus}}\right)^{-1} \nonumber \\
&& \left( \frac{\dot{M}}{1.8 \times 10^{-9} M_{\odot} /yr } \right)^{-0.85} 
\end{eqnarray}
 where $a_i$ is the semi-major axis, $M_i$ is the planetary mass, and $\dot{M}$ is the rate of mass accretion
through the disk (assuming a central star of $1 M_{\odot}$).
The  time dependance of $\tau_{i}$ is driven by the evolution of
the accretion rate,
which is
\begin{equation} 
\dot{M} = \frac{2.5 \times 10^{-8} M_{\odot} yr^{-1} exp(-t/0.63 Myr)}{1 + exp((t-1.7 Myr)/0.15 Myr)} .
\end{equation}
This is chosen to represent a disk that decays exponentially  with time $t$ as it accretes onto the star but then cuts off more
rapidly once photoevaporation losses detach the inner disk from mass resupply by the outer disk \citep{APA14}.

Once a planet has crossed interior to the torque reversal line $a_R$ -- located at the disk surface density maximum \citep{YHH23}, the migration torque is now directed outwards. We 
describe this using a negative timescale. The functional form is again determined by fitting to the simulations
of \cite{YHH23}, and is expressed as a function of position relative to $a_R$, $z=\ln a_i - a_R$. This
amounts to 
\begin{eqnarray}
\left|\tau_{a,i}\right| &=& 6.66 \times 10^{-18} {\rm Myr} \left( \frac{0.1 AU}{a_i}\right)^{1/2} \left( \frac{10 M_{\oplus}}{M_i}\right) \label{taua} \\
&& \left( \frac{\dot{M}}{1.8 \times 10^{-9} M_{\odot}/yr} \right)^{-1.8} 
 e^{-(120.04 + 177.76 z +65.5 z^2 + 8.07 z^3) } \nonumber
\end{eqnarray}

These quantities are furthermore linked by the fact that the transition radius $a_R$ is, itself, a function
of $\dot{M}$ and thus of time, with
\begin{equation}
a_R = 0.062 AU \left[ 1 + 0.229 \ln \left( \frac{\dot{M}}{2.5 \times 10^{-8} M_{\odot}/yr } \right) \right]^{-1}  \label{TTrans} 
\end{equation}
in the case of our default model \footnote{We set the  maximum value of $a_R=1$AU. The exact value is not important as the
migration freezes out long before this occurs.}.
Indeed, if the migration time inward or outward is shorter than the timescale on which $a_R$ is evolving, then the 
planet will track the torque reversal location, making the relevant characteristic timescale $\tau_R=dt/d\ln a_R$.

\begin{figure}
\centering
\includegraphics[width=1.0\linewidth]{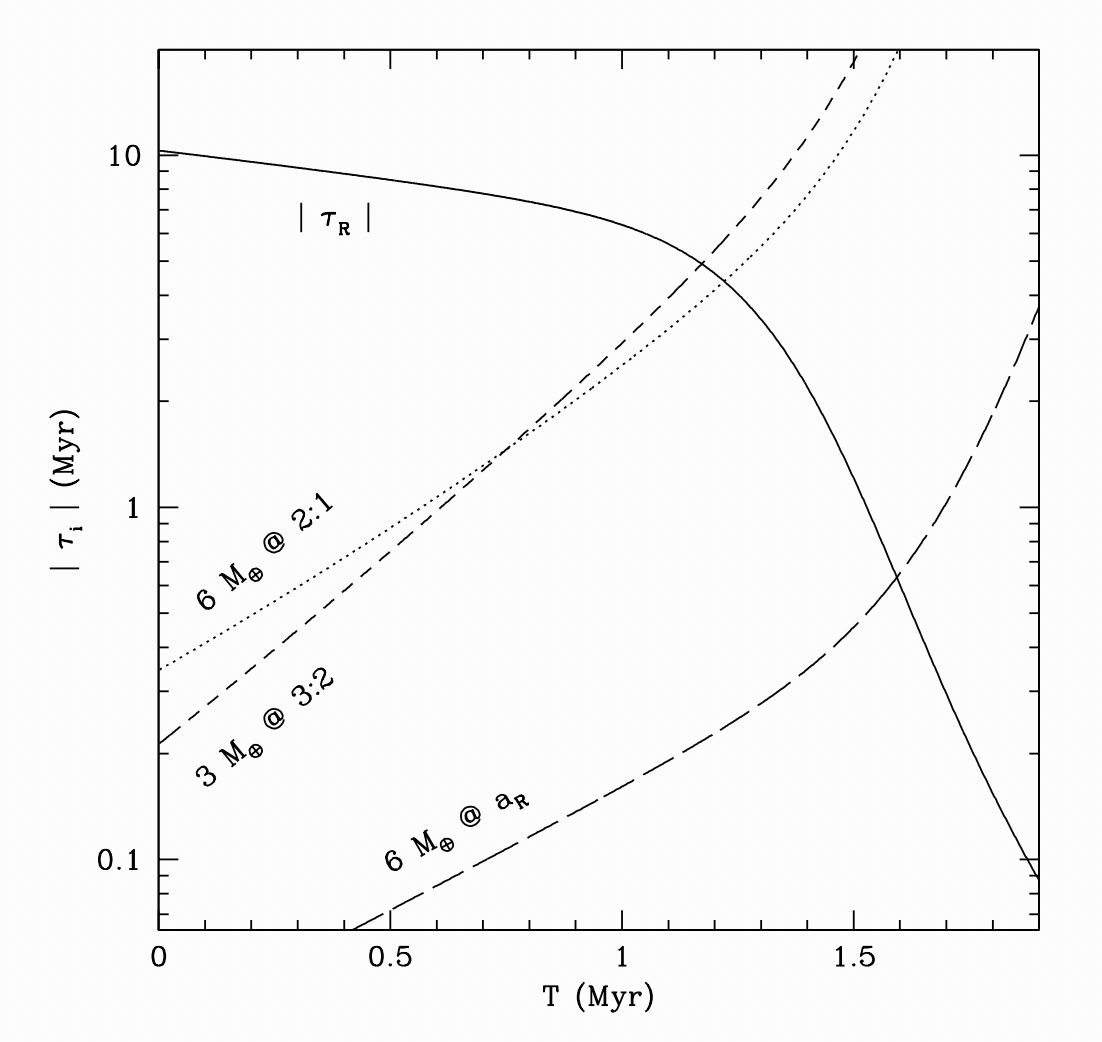}
\caption{The solid curve shows the characteristic timescale for the evolution of $a_R$ as a function of time. We
see that this is initially much longer than the lifetime of the disk, but 
 shortens as time increases, i.e. the torque reversal moves out faster as $\dot{M}$ drops. The dotted
curve represents $\left| \tau_i \right|$ at a location $0.63 a_R$, assuming a mass $6 M_{\oplus}$. This represents the rate
of outward migration for a planet of that mass, trapped in a 2:1 resonance with a planet located at the torque reversal
location. The dashed line represents the same but for a $3 M_{\oplus}$ planet located at $0.763 a_R$, i.e. in a 3:2 resonance.
  Both timescales are shorter than
$\left| \tau_R \right|$  until about 1.2~Myr. After this, the planets cannot keep up with $a_R$ and will freeze out and decouple 
from the disk. The long dashed line indicates the timescale for inward migration, evaluated at $a_R$.  This is shorter than the timescales for the inner planet, indicating that the outer member of the pair will freeze out at a later time.
 \label{TT}}
\end{figure}

For example, if the outer planet of the pair reaches the torque reversal location, then the pair of planets will
both be migrating outwards. The question of whether they converge or diverge will rest on whether the outward
migration of the inner planet is faster or slower than the rate at which $a_R$ recedes because the outer planet is locked
to $a_R$ by the fact that the torques on either side of this transition drive the planet towards it. Two examples of this
 evolution are shown in Figure~\ref{TT}. The solid curve shows the characteristic timescale $\left| \tau_R\right|$ associated with the
outward movement of the torque reversal location as $\dot{M}$ decreases. We also show two cases of $\left| \tau_{a,i} \right|$
interior to this. The dotted line shows the rate of outward migration for a $6 M_{\oplus}$ planet that is the interior
member of a pair locked in 
 the 2:1 mean motion resonance, in which the outer member is  located at $a_R$. At early times, $\left| \tau_{a,i} \right|<
 \left| \tau_R \right|$, so that the planet is able to keep up with the outward migration of the outer member of the pair (which is tied to the torque reversal location $a_R$). Thus, the pair is converging and remains trapped in resonance, even during the outward migration.
  Figure~\ref{TT} shows that $\left| \tau_{a,i} \right|$ increases with time, while the solid curve ($\left| \tau_R \right|$) decreases with time.
 Therefore, when these two curves cross, the inner planet falls behind and the pair starts to diverge, falling out of resonance.

The dashed line shows the case of a $3 M_{\oplus}$ planet locked in a 3:2 resonance with an outer member located at $a_R$. This shows the same characteristic
behaviour, although the divergence starts earlier.  The long dashed curve shows the inward migration timescale for a planet located just exterior to $a_R$. The fact that this curve falls below the others means that the outer member of the pair remains locked to $a_R$ even as the inner member drops out of resonance, although eventually this planet falls behind the retreat of $a_R$ as well.

The
specific value of $a_R$ also depends on the stellar magnetic field and the degree to which it diffuses into the
disk. For a simple dipole field (n=3), equation~(\ref{TTrans}) is equivalent to a magnetic field on the stellar
surface of 300~G. The magnetic fields observed in T~Tauri stars can reach several kG \citep{JK07} but many stars
also show smaller or unobservable fields \citep{DGM11,VAH19}. Furthermore, 
 \cite{YHH23} showed that the results of the disk simulations 
scaled with the enclosed magnetic flux within the planetary radius. Thus, equation~(\ref{TTrans}) also represents a surface field
of $B_0= 1.7$~kG with an effective n=3.5.
Given this ambiguity in trading off $B_0$ and $n$, and the range of potential natal stellar fields, we shall investigate the effects of varying magnetic field by simply scaling the value of $a_R$ in subsequent sections. 

A full description of the interaction between the star and the disk must also account for the fact that there is a true inner boundary
at the edge of the magnetospheric cavity. In most of the cases we discuss here, the planets do not penetrate in far enough to reach
this boundary, and so we will simply consider the magnetospheric cavity to lie at $a_R/3.7$ unless otherwise noted.

 In the following sections, we calculate the evolution of two and three planet systems evolving according to these torques. In
addition to the decay of the semi-major axis, these torques also damp eccentricities. Resonant interactions between neighbouring planets lead to a coupled evolution, including eccentricity excitation and resonant lock. The equations that describe this evolution are given
in Appendix~\ref{Eqns}.

\section{Converging versus Diverging}
\label{Times}

When two planets converge during migration, they enter a state of resonant lock, in which the rate of change
of orbital frequencies are equal, i.e. $\dot{n_1}/n_1 = \dot{n_2}/n_2$. If we simultaneously impose the condition
that the libration amplitude of the resonant angle is small,  the eccentricities reach equilibrium values given by
\citep{TP19}, in the ratio
\begin{equation}
\left( \frac{e_2}{e_1}\right)^2 = \frac{1}{4} \left( \frac{m_1}{\alpha_{12} m_2} \right)^2  \left( \frac{f_2''}{f_1} \right)^2 \label{EqEcc}
\end{equation}
where we have chosen the q=1 resonance for this example and the various quantities are defined  using the
conventions given  in appendix~\ref{Eqns}. These conditions also imply specific values
for the equilibrium eccentricities. The full expressions are given in \cite{TP19}, but the most important element
is that
\begin{equation}
e_1^2 \propto \left(\frac{t_{e,1}}{t_{a,2}} - \frac{t_{e,1}}{t_{a,1}} \right) \label{EqT}
\end{equation}
In the traditional case where both planets are migrating inwards, this quantity is positive as long as $t_{a,1}>t_{a,2}$, i.e. convergent
migration yields resonant lock. In our model, the torques can reverse sign, which can be modelled as simply reversing the sign of
the relevant timescale.

Thus, when the innermost planet first crosses the torque reversal location, the value of $t_{a,1}$ will change sign. This will cause the
pair to converge even more strongly, so that there will still be an equilibrium, although the values of the equilibrium eccentricities will change.  This can occur even if the resonant lock does not change.
However, once the second planet crosses the torque reversal location, then both planets are moving outwards and the
 question of whether they are converging or diverging is once again a matter of their relative motion. The equilibrium will fail if the quantity in equation~(\ref{EqEcc}) becomes
negative -- thereby breaking the resonant lock. The system evolution retains the same character if one of the planets is tracking
the location of the torque reversal -- the relative $t_{a}$ is then simply dictated by $\tau_R$ -- the evolution of the torque reversal.

Extending the analysis to three planet systems yields qualitatively similar behavior, with locked resonant systems exhibiting equilibrium eccentricities in fixed proportions, but also demonstrating failure of these equilibria once the planets start to move outwards again. 

\section{Two Planet Systems}
\label{Dos}

\begin{figure}
\centering
\includegraphics[width=1.0\linewidth]{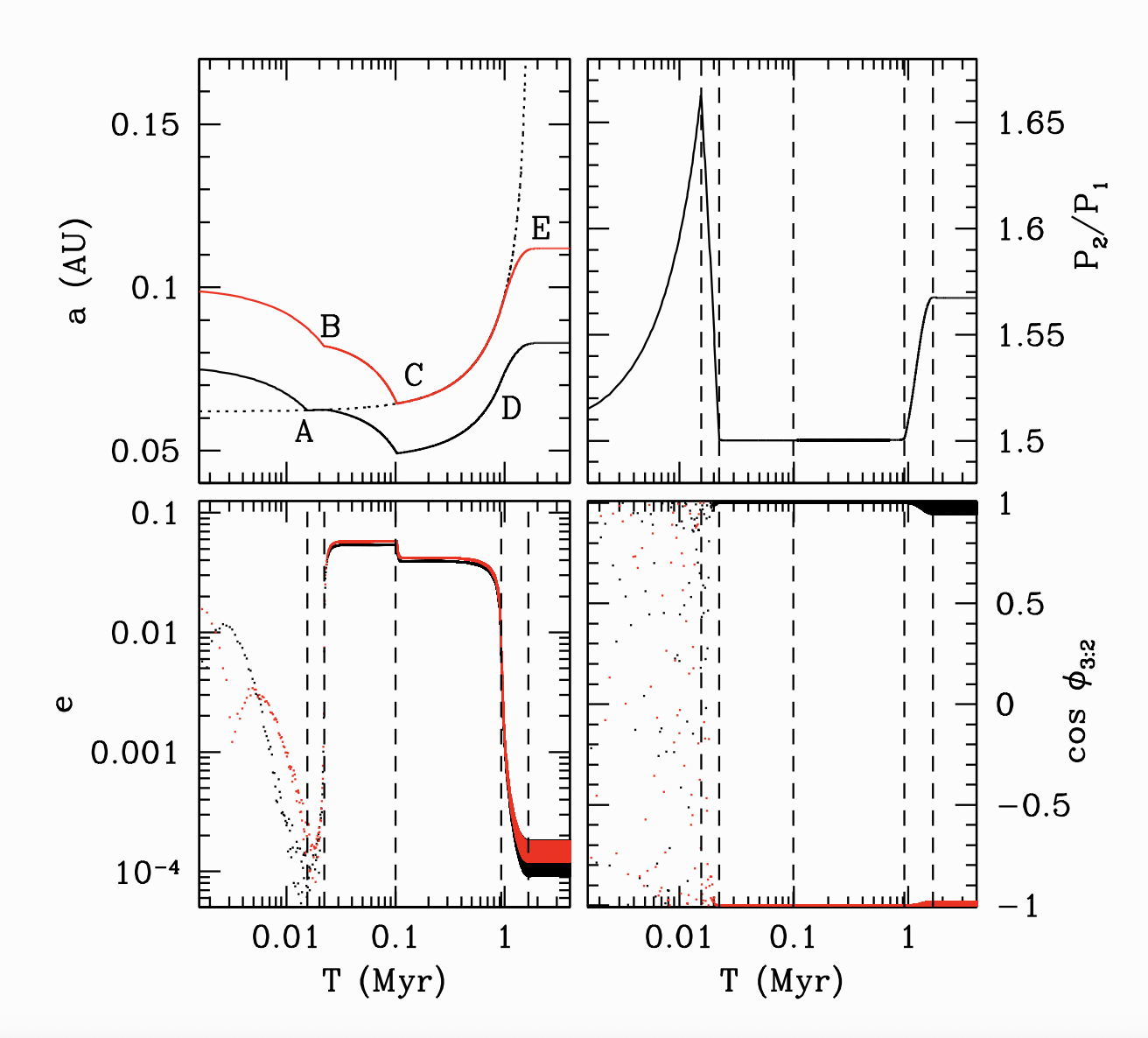}
\caption{The upper left panel shows the evolution of semi-major axis for a pair of $3 M_{\oplus}$ planets migrating due to
the gaseous torque model described in the text. The dotted curve shows how the location of the torque reversal evolves with
time. Important transitions in the dynamical state of the system are labelled as A--E, and are shown as vertical dashed lines in the 
other three panels. In the upper right, we see the evolution of the period ratio for the pair. In the lower left we show the evolution
of the eccentricity for each planet (red curve indicating the outer planet and black curve the inner planet). In the bottom right, we see
the evolution of the cosines of the resonant angles $\phi_{3:2}$, namely 
 $3 \lambda_2 - 2 \lambda_1 - \bar{\omega}_1$ (black) and $3 \lambda_2 - 2 \lambda_1 - \bar{\omega}_2$ (red).
 \label{Pair1}}
\end{figure}

Figure~\ref{Pair1} illustrates the basic evolution for a two planet system, containing two $3 M_{\oplus}$ planets. 
  The initial semi-major axes are 0.076~AU and 0.1~AU, with initial eccentricities of 0.01. The initial mean longitudes and longitudes of periastron were chosen randomly and the relevant resonant angles constructed therefrom. The equations in Appendix~\ref{Eqns} are integrated using the adaptive stepsize integrator {\it odeint} from \cite{NumRec}.
The black curve in the upper left plot
illustrates the evolution of the semi-major axis of the inner planet, and the red curve the evolution of the outer planet. The dotted line indicates the evolution of the location of the torque reversal location $a_R$. The upper right
panel shows the period ratio of the two planets as a function of time. In the lower left, we show the evolution of the planetary eccentricities,
and, in the lower right, the panel shows how the resonant angle for the inner and outer 3:2 resonances change with the planetary and disk evolution.

The first incident of note -- denoted point A -- is when the inner planet reaches
the torque reversal location $a_R$. At this point the inward migration of the inner planet stops and the planet semi-major axis starts to track the evolution of $a_R$, which is slowly moving outwards. Up until this point, the period ratio is increasing (the planets are diverging) because the migration time decreases as planets get closer to the star. However, with the halting of the inner planet migration at A, the planets
begin to converge because the outer
planet, shown as the red curve, continues to migrate inwards, and the two planets enter the 3:2 resonance at B. The net relative motion is convergent and so the pair become locked in a resonant configuration. As a result, the eccentricities rise from almost zero to finite values dictated by the equilibrium defined by equation~(\ref{EqEcc}).

The torques on the outer planet now drive the pair inwards, and the inner planet is pushed interior to the torque reversal. Eventually the
outer planet reaches the torque reversal (at time denoted as C). The planets remain in equilibrium (because the torques on the inner planet are pushing it outwards, so the pair is still converging) but 
 the equilibrium eccentricities change because the sign of
the torques change  which changes the equilibrium condition in equation~(\ref{EqT}). After this, the pair starts to migrate outwards but remains in resonance as long as the rate of outward migration of the inner planet is  faster than the evolution of the torque reversal. This continues for a long time, but eventually, as the torques weaken and the outward motion of $a_R$ accelerates, the pair starts to diverge, at epoch D. The libration in
the 3:2 resonance  begins to increase, as does the period ratio, and the eccentricities start to damp away as their resonant excitation
weakens.
Eventually the migration freezes out completely (at E), leaving the system still in resonance with a finite libration amplitude and an offset 
relative to strict commensurability.

This is the fundamental result that emerges from these calculations -- in the disk models of \cite{YHH23}, planets are not trapped at the true inner edge of the disk, but rather at the location of the torque reversal. As the disk evolves and the surface density drops, this location moves outwards, resulting in an outward migration of the planets at, and interior to, the torque reversal. In many cases, this leads to divergent migration, thereby breaking the resonant lock and causing divergence of the period ratios away from the initial commensurabilities.

\begin{figure}
\centering
\includegraphics[width=1.0\linewidth]{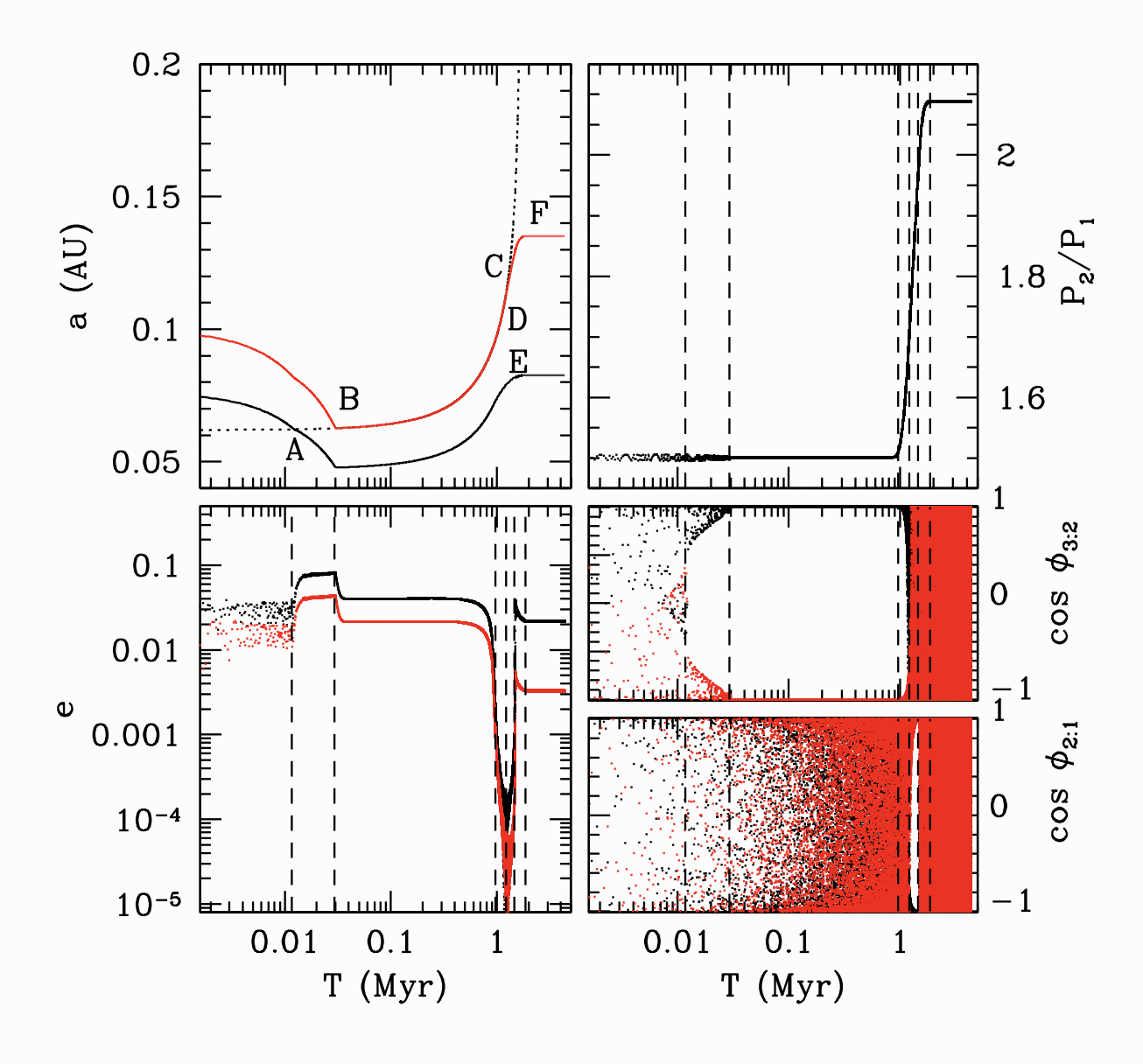}
\caption{The upper left panel shows the evolution of semi-major axis for a pair of migrating planets where the
outer planet is $6 M_{\oplus}$ and the inner planet is $3 M_{\oplus}$. 
The dotted curve shows how $a_R$ evolves with
time. Important transitions in the dynamical state of the system are labelled as A--F, and are shown as vertical dashed lines in the 
other three panels. In the upper right, we see the evolution of the period ratio for the pair. In the lower left we show the evolution
of the eccentricity for each planet (red curve indicating the outer planet and black curve the inner planet). In the bottom right, we 
have two panels to track the evolution  of the cosines of the resonant angles for the two important first order resonances that are encountered.
The one designated as $\phi_{3:2}$ shows the angles $3 \lambda_2 - 2 \lambda_1 -  \bar{\omega}_1$ (black) and
$3 \lambda_2 - 2 \lambda_1 - \bar{\omega}_2$ (red). The one designated as $\phi_{2:1}$ shows 
the evolution of the two resonant angles $2 \lambda_2 - \lambda_1 - \bar{\omega}_1$ (black) and $2 \lambda_2 - \lambda_1 - \bar{\omega}_2$ (red).  \label{Pair2}}
\end{figure}

The planetary masses also play a role in determining the outcome of the evolution, because more massive planets remain coupled to the gas disk for longer, and lower mass planets freeze out earlier. To illustrate this, 
Figure~\ref{Pair2} shows the case of a two planet system wherein the outer planet ($6 M_{\oplus}$) is more massive than the inner one ($3 M_{\oplus}$).  The initial semi-major axes, eccentricities and longitudes are the same as in Figure~\ref{Pair1} and 
the initial evolution is qualitatively similar, except that the stronger torques on the outer planet means that the inner planet
barely halts at the torque reversal before being captured in resonance and pushed further inwards. The eventual divergence from the 3:2 resonance (at D) is also qualitatively similar. However, the outer planet remains coupled to the gas disk for longer by virtue of its larger mass and so migrates far enough that the system passes through the 2:1 resonance (shown at D). This resonant encounter is now divergent, so there is no capture (the system leaves the 2:1 resonance at E)  but there is a transient excitation of the eccentricities that occurs late enough to leave small non-zero
eccentricities ($\sim 0.01$) after the system has frozen out (at F).

\begin{figure}
\centering
\includegraphics[width=1.0\linewidth]{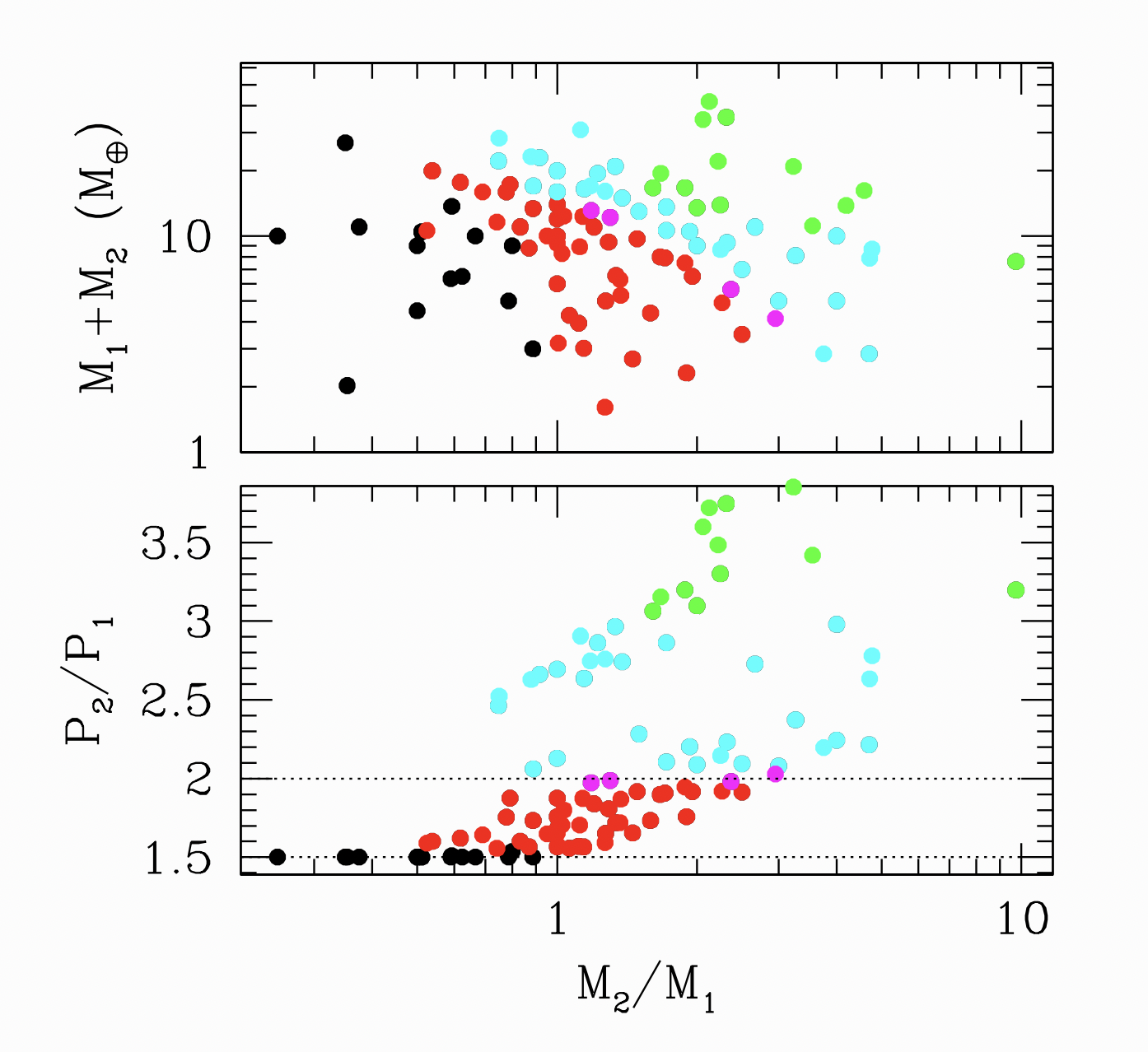}
\caption{The lower panel shows the final period ratio of a planetary pair as a function of the mass ratio $M_2/M_1$.
The results are color coded according to the final value, with black signifying $P_2/P_1<1.55$, red (1.55--1.95),
magenta (1.95--2.05), cyan (2.05--3) and green for $P_2/P_1>3$. In the upper panel, these colors indicate the final
states of systems as a function of mass ratio and total system mass $M_1+M_2$. The dotted lines in the lower
panel indicate the first order commensurabilities (1+q)/q for q=2 and q=1. \label{Mrat2}}
\end{figure}

The overall amount of migration is driven largely by the planetary masses, as they determine how long the planet remains
coupled to the disk and how rapidly the planets migrate with respect to one another. Figure~\ref{Mrat2} shows the effect
of both planetary mass ratio and total mass in the overall evolution of a range of planet pairs.
 Each pair starts with the same semi-major axes and eccentricities as in Figure~\ref{Pair1}, while the initial mean longitudes and
longitudes of periastron are chosen
randomly. We see that 
low mass ratios ($M_2/M_1 <0.6$) leave pairs trapped in the original 3:2 resonance, whereas larger mass ratios lead to divergence and a wide range
of final period ratios, including some in the range of the wider 2:1 resonance. This is a consequence of the fact that the more massive planets remain coupled to the gas for longer. Thus,
if the inner planet is the more massive one, the pair will remain in the convergent state for the entirety of the disk evolution. However,
if the outer planet is the more massive one, then the inner planet will freeze out first and fall behind the outer planet's migration. This leads to a diverging configuration and a final period ratio that increases as $M_2/M_1$ does. The degree of divergence also depends on the
total planet mass, with more massive planetary systems leading to large final mass ratios (because they remain coupled to the disk
for longer).

\subsection{Dependance on model parameters}

\begin{figure}
\centering
\includegraphics[width=1.0\linewidth]{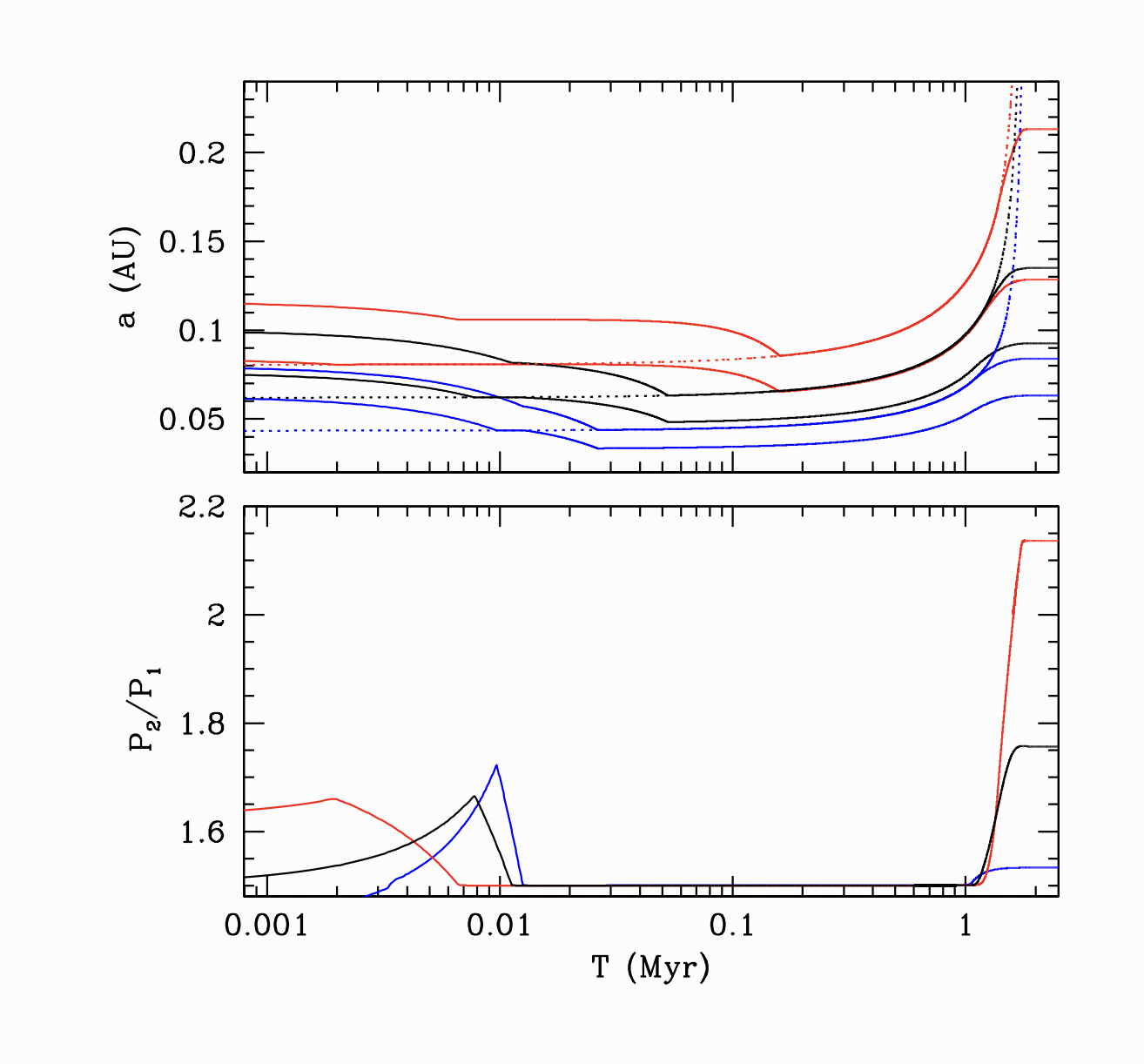}
\caption{The upper curves show the evolution of the semi-major axes of  pairs of $6 M_{\oplus}$ planets 
evolving in disks with different torque reversal locations ($a_R$). The black curve corresponds to
the model given by equation~(\ref{TTrans}). The red curves correspond to $a_R$ scaled up by a factor 1.3, and
the blue curves have $a_R$ scaled down by 0.7. The dotted lines show the evolution of the respective
$a_R$. The lower panel shows the evolution of the corresponding period ratios. \label{Pairt2}}
\end{figure}

The evolution of these planetary pairs is driven by the torques due to the gas disk, so they will be influenced by the
parameters of the gaseous disk, and also by the strength of the stellar magnetic field -- which determines the location
of the torque reversal that drives the outward motion and divergence. Figure~\ref{Pairt2} illustrates this.

Figure~\ref{Pairt2} shows the evolution of a pair of $6 M_{\oplus}$ planets in three different scenarios. In each
case, the accretion disk evolution is the same, but we have scaled the form of $a_R$, given in equation~(\ref{TTrans}),
by a constant factor. As discussed above, this mimics variations in either the stellar magnetic field strength, or the degree
of field compression by the inflow ($n$). Thus the black curve in Figure~\ref{Pairt2} shows the 
 default case, while the blue curve shows the case
where $a_T$ is a factor 0.7 smaller (corresponding to a weaker magnetospheric influence) and the red case represents the case where
$a_T$ is a factor 1.3 larger (corresponding to a stronger magnetospheric influence).  In each case, the initial semi-major axes are chosen such that all planets start exterior to $a_R$. The blue curves start with semi-major axes 0.063 and 0.08~AU, the black curves with 0.076 and 0.1~AU, and the red curves with 0.084~AU and 0.116~AU. Initial eccentricities are still 0.01 and longitudes are chosen at random.

We see that more distant torque reversals result in an earlier entry into
resonance, and a slightly later departure from commensurability. The end result is a more widely
separated pair of planets. Given that we expect a range of stellar magnetic fields amongst exoplanet
host stars, this will also result in some variation of the planetary spacings.

\begin{figure}
\centering
\includegraphics[width=1.0\linewidth]{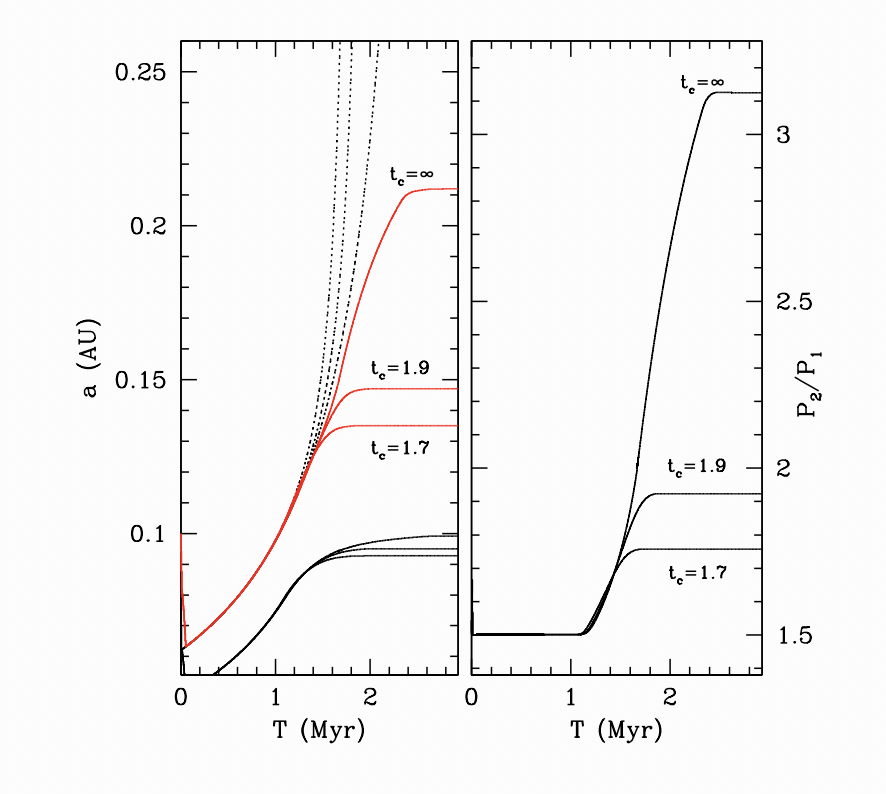}
\caption{The curves show the evolution of a pair of $6 M_{\oplus}$ planets in the same disk, with the
same initial conditions, but with three different values for the time ($t_c$) at which the accretion rate starts
to rapidly drop off. In the left panel, the three black curves represent the inner planet and the three
red curves the outer planet. The right hand panel shows the evolution of the respective period ratios.
 \label{Pair4}}
\end{figure}

The final separations also depend on the model for the evolution of the accretion rate and surface
density of the disk. Our default model assumes an evolution characterised by two exponentials, chosen
to mimic the slow decay of a viscously evolving disk that accelerates once the inner disk has been decoupled
from the outer disk by photoevaporation. Figure~\ref{Pair4} shows how the evolution of a pair of $6 M_{\oplus}$
planets changes depending on the choice of the faster exponential cutoff.  Initial conditions for each case are the
same as those for Figure~\ref{Pair1}. We show the evolution where the cutoff
rolls over after 1.7~Myr, 1.9~Myr or not at all (i.e. the decay is characterised by only one exponential -- the slower
one). We see that later cutoffs lead to more divergence.

\section{Three Planet Systems}
\label{Tres}

In the previous section, we have demonstrated that the presence of a transition region and torque reversal in the inner disk can
cause a significant dynamical shake-up in planetary systems, potentially explaining the low frequency of observed resonant pairs.
The sample of planets discovered by the Kepler satellite contains many systems with multiplicity higher than two. In this section we will examine
the effects of a torque reversal on three planet systems.

The addition of a third planet adds some complexity to the system, as the planet in the middle now experiences gravitational interactions
from both the inner and outer companions. As a result, this planet may be evolving outwards with the torque reversal -- and diverging from the inner planet -- but
then experiences torques from the third, outer, planet, which is still being driven inwards. This influence can counter the evolution discussed in the previous section, and so we must consider the parameters associated with this competition.

\begin{figure}
\centering
\includegraphics[width=1.0\linewidth]{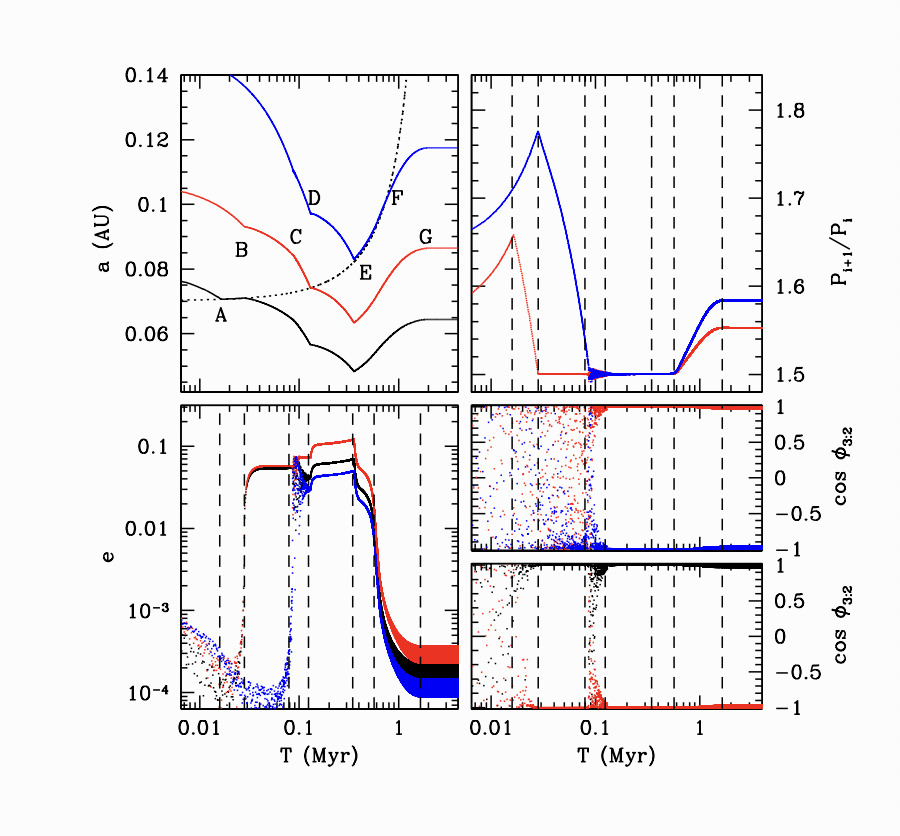}
\caption{ Here we show a three planet system, containing three planets, each with mass $3 M_{\oplus}$. The evolution
of the inner planet is shown in black, the middle planet in red, and the outer planet in blue. 
In the upper right panel, the red curve refers to period ratio of the inner pair, while the blue curve indicates the period ratio of the
outer pair. In the lower right, we show the behaviour of the  cosines of the resonant angles for  the 3:2 resonance for both
 the outer pair (red and blue) and inner pair (red and black).
The labels in the upper left indicate important epochs in the evolution of the system, identified as vertical dashed lines in
the other panels. \label{Trip1}}
\end{figure}

Figure~\ref{Trip1} shows the simple extension of our model system shown in Figure~\ref{Pair1}, in which we simply add a 
third planet, also of 
 mass  $3 M_{\oplus}$.  The initial semi-major axes are now 0.08, 0.107 and 0.149~AU, 
 chosen so that both pairs begin between the 3:2 and 2:1 commensurabilities.  Eccentricities and longitudes are chosen as before.
The initial evolution is divergent,
because the rate of migration is faster closer to the star.  The first planet encounters the torque reversal at A, and so the inner
pair begins to converge, trapping into the 3:2 resonance at B. This excites the eccentricities of the inner pair to equilibrium values, but the outer planet remains almost circular as it continues to migrate inwards. The resonant interaction slows the inward migration of the  middle
planet and the outer pair encounters the 3:2 resonance at C. This also briefly disturbs the resonance lock of the inner pair, but both pairs soon settle into resonance again when the middle planet reaches $a_R$ (at D), along with a new set of equilibrium eccentricity ratios. The outer planet continues to migrate inwards, pushing both the inner and middle planets interior to $a_R$, until it eventually encounters the torque reversal as well (at E). At this point we have
three planets in a resonant chain of 3:2 resonances, moving outwards with the outer planet tracking $a_R$. As we have noted above, this will continue until the evolution of $a_R$ accelerates and the planets start to diverge (at F).  The divergence is not rapid, and the planets eventually freeze out (at G) still in resonance, with a small libration, and moderate separations from commensurability.

The end result is a trio of planets, with the inner pair having period ratio $P_2/P_1=1.54$, and the outer pair with a period
ratio $P_3/P_2=1.57$ -- similar to what one would expect by simply combining the two pairs independently.

\begin{figure}
\centering
\includegraphics[width=1.0\linewidth]{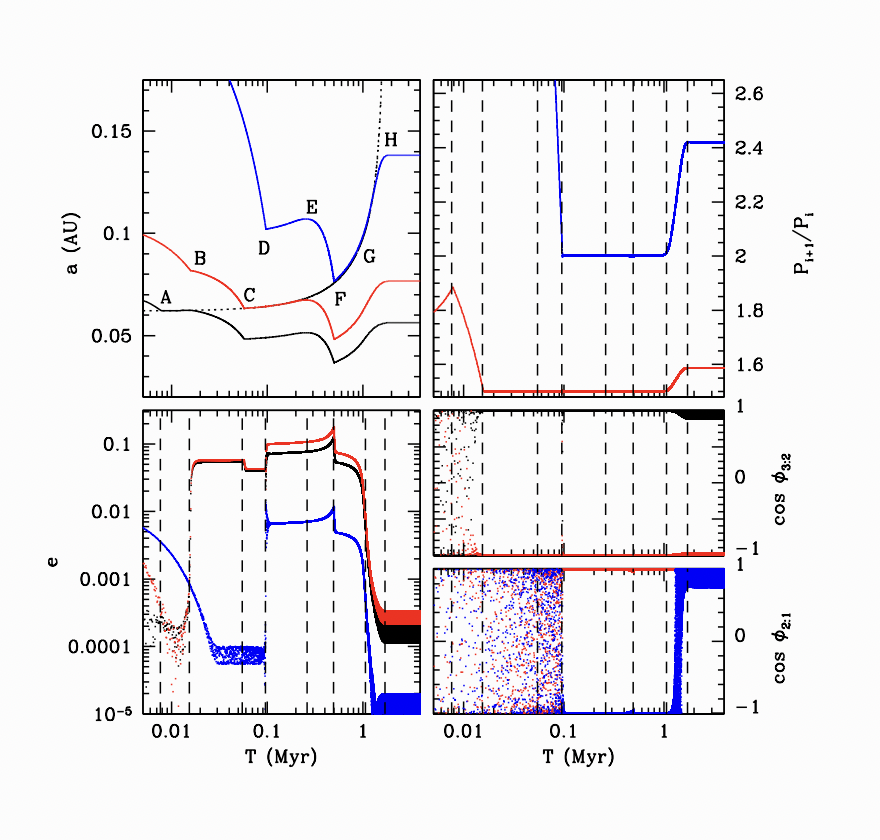}
\caption{ Here we show a three planet system, containing three planets, each with mass $6 M_{\oplus}$. The evolution
of the inner planet is shown in black, the middle planet in red, and the outer planet in blue. 
In the upper right panel, the red curve refers to period ratio of the inner pair, while the blue curve indicates the period ratio of the
outer pair. In the lower right, we show the behaviour of the resonant angles for the different first order resonances at play.
The upper panel in this quadrant shows the evolution of the cosine of the resonant angle for 
the 3:2 resonance for 
 the inner pair (red and black). The bottom  panel show the corresponding evolution
 for the 2:1 resonance between the outer pair of planets.
The labels in the upper left indicate important epochs in the evolution of the system, identified as vertical dashed lines in
the other panels. \label{Trip2}}
\end{figure}

Figure~\ref{Trip2} shows the effect of  more massive planets and wider orbits.
All the masses are scaled  up by a factor of 2 -- i.e. a system containing three planets
of mass $6 M_{\oplus}$. Initial semi-major axes are 0.076, 0.107 and 0.223~AU, so that the outer pair starts wide of
the 2:1 resonance. 
 The inner pair evolves in the same manner as before -- converging after the inner planet
reaches the torque reversal (A), locking into the 3:2 resonance (B) and finding a new equilibrium when the middle
planet reaches the torque reversal (C). The outer pair
locks into the 2:1 resonance, at D. However, the larger masses mean that the outward torque is stronger and
the mutual 
gravitational  interactions are strong enough to prevent the outer planet
from forcing its way inward to $a_R$. Thus, the system evolves for a period of time with the middle
planet tracking $a_R$ and maintaining an inner 3:2 and outer 2:1 resonance,  with the outer planet being
driven outwards despite not having reached $a_R$.

Eventually, as the outwards evolution of $a_R$ accelerates, the middle planet detaches from the torque
reversal and the outer planet resumes its inward migration (E), eventually reaching $a_R$ at F. At this point
the evolution is similar again, with the planets starting to diverge from strict commensurability as the outer planet
decouples from the torque reversal (G) and eventually freezing out at H. One noteworthy difference here though, is the
behaviour of the outer pair, wherein the outer 2:1 resonant angle reverses its libration orientation.  This occurs when
the period ratio diverges beyond $P_3/P_2=2.25$, at which point the stable libration equilibrium switches in angle
by $\pi$ (as discussed in appendix~\ref{Eqns}).

\subsection{Observational Comparison}

These calculations demonstrate that the evolution of systems of multiple planets in the protoplanetary disk background 
calculated by \cite{YHH23} can produce divergence from resonance and result in final period ratios well away from the
commensurabilities expected in a naive application of the resonant locking theory. 

In a perfect scenario, we should be able to model the divergence of specific planetary systems with well-defined masses,
although the stellar magnetic field value and the evolution of the protoplanetary disk remain quantities that need to be
modelled and so can introduce
some variations.  Of course, the observational reality is less well constrained -- most of the known multiplanet systems are detected
via the transit method, and most of the planetary masses are small enough to prove challenging to measure with the radial
velocity method.

Nevertheless, it is instructive to compare our existing model with the few known planetary systems where the planetary masses
are well measured. As a comparison for our model, we require systems of three planets, orbiting approximately Solar-mass stars, and with
well measured masses. We restrict ourselves solely to triple planet systems, since the presence of neighbors does influence the
degree of divergence.

\begin{figure}
\centering
\includegraphics[width=1.0\linewidth]{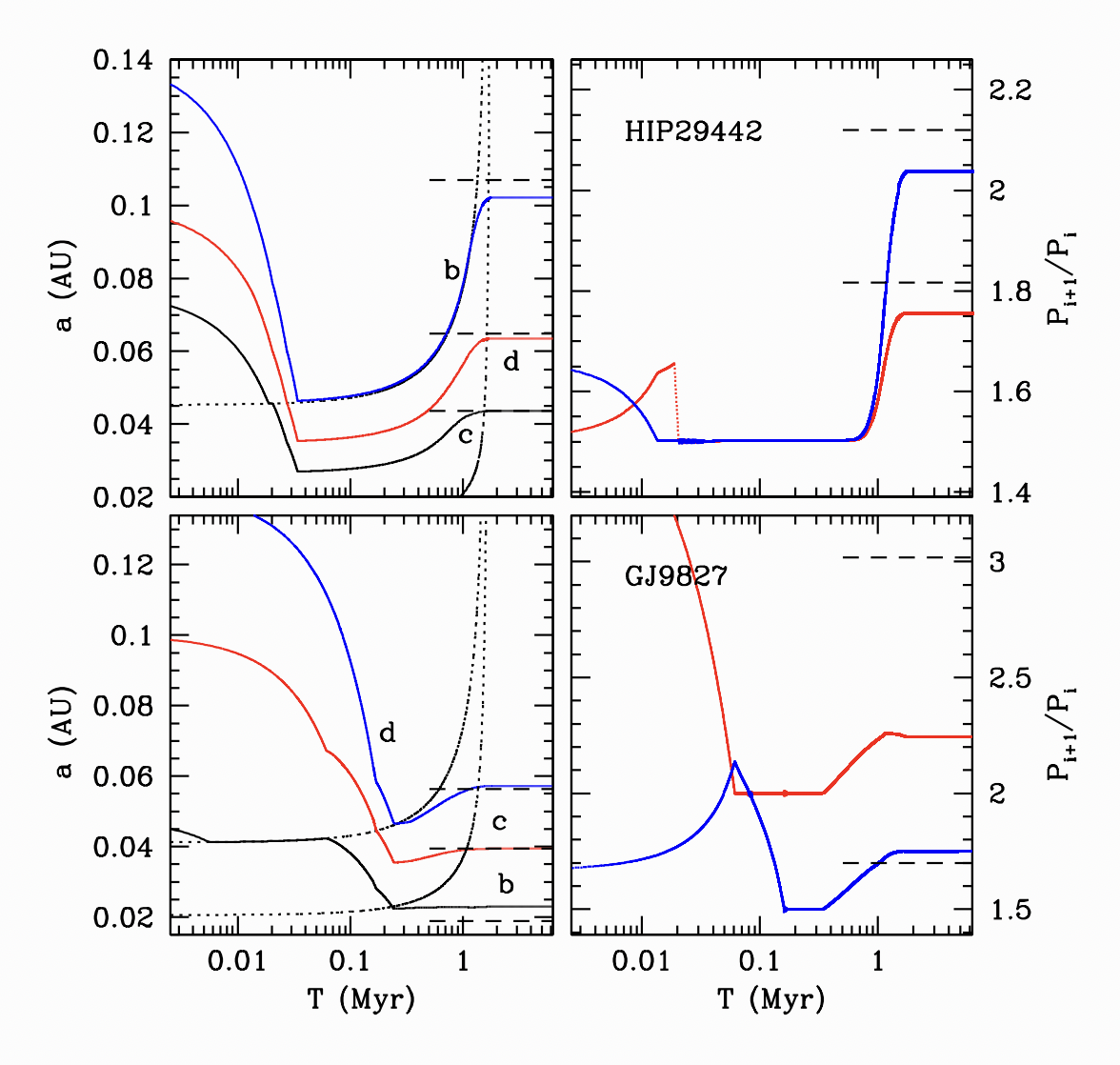}
\caption{ The curves  in the upper two panels show the evolution of a planetary system modelled after the HIP29442 planetary system, assuming
our standard disk and a torque reversal scaled by a factor 0.73. The horizontal dashed lines in the left hand panel indicate
the observed final semi-major axes and the horizontal dashed lines in the upper right panel indicate the observed final period ratios.
The lower panels show the equivalent evolutions for the GJ9827 system, in which the torque reversal is scaled by 0.55. In both
panels, the upper dotted line indicates the torque reversal $a_R$, while the lower dotted line indicates the edge of the magnetospheric cavity, where the torques go to zero. \label{TripEx}}
\end{figure}

 Figure~\ref{TripEx} shows a comparison for the two planetary systems that best match our criteria. In the top panel we show a 
model for the planetary system around the star HIP29442. This
system has a host star that it is close to solar mass ($0.89 M_{\odot}$), and has exactly three
planets with well-constrained masses (errors $<1 M_{\oplus}$).  In the bottom panel, we show a comparison to the star GJ9827.
The host star in this system is $0.62 M_{\odot}$, which is probably too low to be strictly applicable to our model, but is the next closest match and still instructive.
Other systems with well measured masses either orbit
M dwarfs or have either more or less than three planets.

The  planetary masses for HIP29442 measured by \cite{Dam23} are $4.5 \pm 0.3 M_{\oplus}$, $5.1 \pm 0.4 M_{\oplus}$ and $9.6 \pm 0.8 M_{\oplus}$
for the three planets with orbital periods of 3.538~days (c), 6.429~days (d) and 13.63~days (b).  To model this system
we choose the scaling of $a_R$ 
so that the innermost planet freezes out at approximately the correct distance. The initial semi-major axes are 0.076, 0.1 and 0.14~AU, with initial eccentricities of 0.01 and random initial longitudes.
  The resulting evolution shows the system enters a 3:2+3:2 resonant chain early
and then diverges to final period ratios that are similar to those observed, although slightly smaller. In particular, we note that the outer pair diverged from interior of the 2:1 resonance to outside it. 

 The masses for the
GJ9827 planets \citep{Pass24} are $4.3 \pm 0.3 M_{\oplus}$, $1.9 \pm 0.4 M_{\oplus}$ and $3.0 \pm 0.6 M_{\oplus}$, corresponding to 
 orbital periods of 1.209~days (b), 3.648~days (c) and 6.202~days (d).  In this case, we scaled $a_R$ so that the middle planet
freezes out at the correct distance, and scaled the accretion rate down by a factor of 0.7 to account for the lower stellar mass.
 The initial conditions are the same as for HIP29442 except that the inner planet starts with semi-major axis 0.048~AU so that it encounters $a_R$ early -- since the observed inner planet is very close to the host star.
 With
these adjustments
 our migration model is able to match the outer pair quite well, but the inner planet does not migrate in as far as desired.
We have also moved the edge of the magnetospheric cavity out to a value of $a_R/2$. This halts the migration of the
inward planet once it crosses into the cavity, but the inward migration is still halted by the reversal of the outermost planet when it hits $a_R$.   Thus, the inner pair mass ratio falls somewhat short of the observed value, but does reproduce the qualitative features of the system in that the inner pair is wider than the 2:1 resonance while the outer pair lies between 3:2 and 2:1.

Of course, most planetary systems do not have masses measured to this accuracy, but our principal goal is not to constrain
the histories of individual planets, but to understand the features of the underlying population of planetary systems.
 The most representative -- in the sense of the most homogenously selected -- sample of low mass planets is that derived from the Kepler mission. Thus, we wish to compare our model to the corresponding sample of planetary systems detected with Kepler.

\subsection{Kepler Analog Systems}

\label{KepAn}

As we can see from Figure~\ref{Mrat2}, the degree of divergence of a planetary pair away from resonance is partially
regulated by the mass ratio. 
Thus, to
survey a realistic range of masses, we choose to simulate all mass combinations detected amongst Kepler triple systems.
 The Kepler results provide radii, not masses, so
we must use an empirical mass-radius relation to convert each set of radii to masses. We use the results
of \cite{CK18}, which provides mass estimates for 7000 KOI, based on the model outlined in \cite{CK17}. In particular, 
this model includes a spread in the
expected masses at fixed radius, in accordance with the observational evidence. However, the possible spread
in masses from planet to planet is quite large, and selecting three masses independently will give a large variation
in mass ratios -- inconsistent with the empirical peas-in-a-pod phenomenon \citep{WMP18}, which states that the
degree of uniformity in planetary properties within a given system is smaller than that within the overall exoplanet
population as a whole.

As a result, we choose the mass of the middle planet from the results of \cite{CK18} and then scale the inner and
outer planets relative to this using the deterministic component of the mass-radius relation in \cite{CK17}. This
allows for some variation in the overall masses, but keeps the internal mass variation consistent with the radius
variations. Using these masses we evolve forward  realisations of the 107 Kepler planetary systems that
contain three confirmed  planets, none of which have a radius $>6 R_{\oplus}$ (our torque model is not applicable
to planets of mass $> 30 M_{\oplus}$ \citep{YHH23}, which will open gaps in the protoplanetary disk).  We wish to focus
on the effects of the outspiral and detachment from the disk at late times, so we assume the planets occupy
an initially compact configuration, starting the planets at 0.076~AU, 0.107~AU and 0.158~AU, so that the systems lie between
the 3:2 and 2:1 resonances. Our model assumes a $1 M_{\odot}$ planetary host star, as 85\% of the 107 Kepler systems to which we compare orbit FGK stars.

\begin{figure}
\centering
\includegraphics[width=1.0\linewidth]{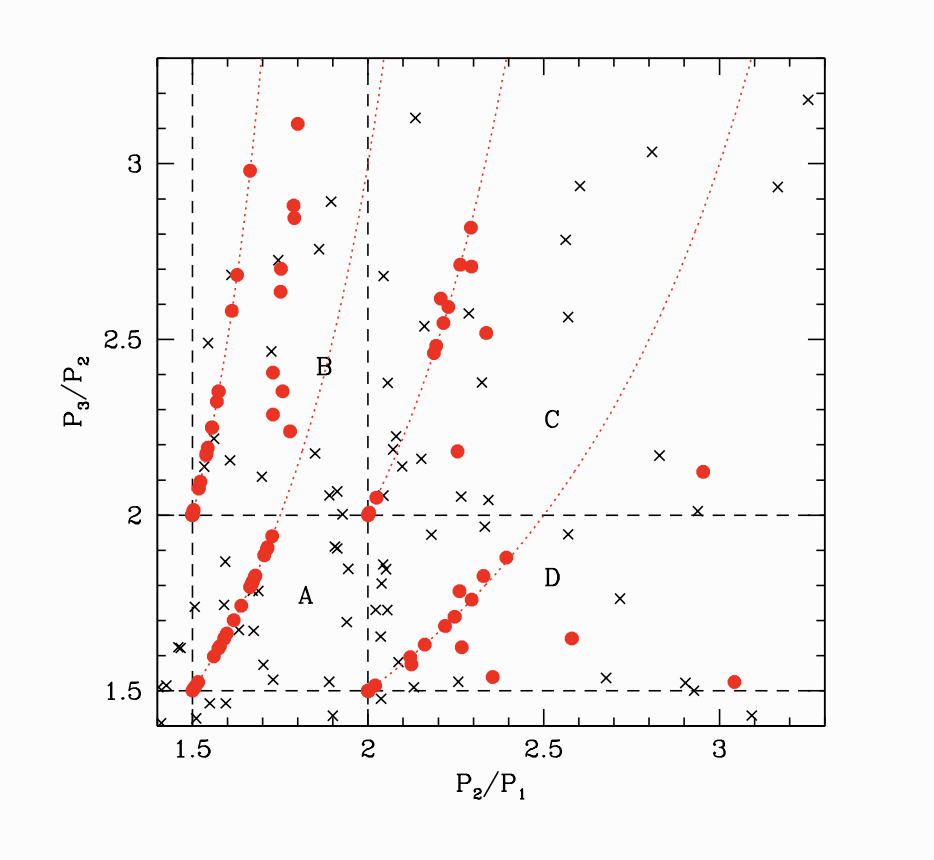}
\caption{ The solid  red points represent the period ratios of inner (horizontal axis) and outer (vertical axis) pairs
from our simulations of the set of Kepler triple analoges.  The red dotted lines indicate the three-body relationship
derived from eliminating the longitude of periastron of the middle planet.
 The crosses correspond to observed Kepler triples, and the dashed lines indicate the
3:2 and 2:1 commensurabilities. The labelling of the four quadrants A--D are discussed in the
text. \label{Mix2}}
\end{figure}

The results of this parameter scan is shown in Figure~\ref{Mix2}. The crosses show the parameters of
the known Kepler three planet systems -- with inner pair period ratio on the horizontal axis and
outer pair period ratio on the vertical axis. The solid  red points show the results of the model calculations.
We have divided this parameter space into four quadrants based on where the period ratios lie relative
to the first order resonances (q+1:q). Systems in quadrant A have both pairs between the q=1 and q=2 first
order resonances. Systems in quadrant B have a compact inner pair (between q=1 and q=2) and a
wider outer pair (wider than q=1). Quadrant C has both pairs wider than the q=1 resonance and
quadrant D has a wider inner pair (outside q=1) and a compact outer pair (between q=2 and q=1).

We see that our model is capable of producing systems that lie in all four quadrants, as well as a handful
that cluster near the resonant chains. This variation is achieved solely as a function of the different masses
-- all of the red points are generated for the same disk evolution model, initial spatial configurations and $a_R$ history. This
demonstrates that the mass variations in the Kepler systems are capable of producing a substantial
fraction of the overall  period ratio variations.  The red points in each quadrant also show an ordered structure in the immediate vicinity of each of the resonant intersections. These are the result of the fact that a resonant chain of two mean motion resonances also results in a three-body resonance, whose expression is derived by eliminating the  longitude of periastron of the middle planet between the two resonances for which it is present \citep{Aks88}. These are shown as red dotted lines in Figure~\ref{Mix2}. It should be noted that our equations describe only the first order two-body resonant interactions and so do not contain any zeroth order contributions to the three-body resonance. Nevertheless, the diverging chains preserve the three body relationship well away from the two-body commensurabilities.

\begin{figure}
\centering
\includegraphics[width=1.0\linewidth]{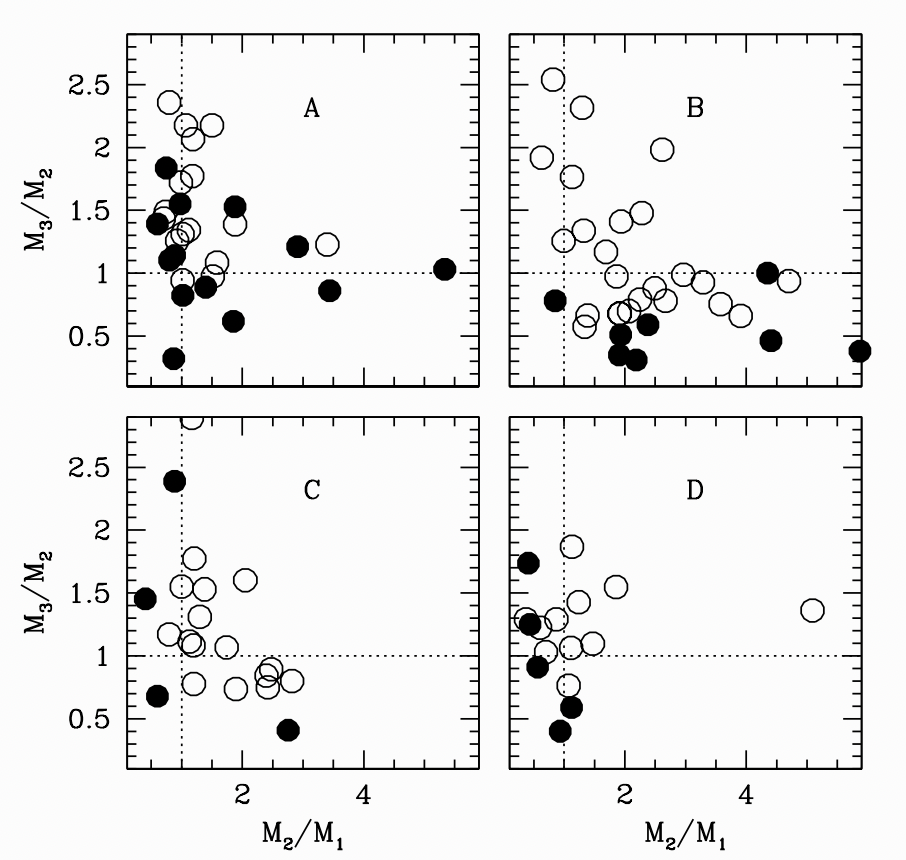}
\caption{ The four panels show the planetary mass ratios for model three planet systems that are found
in each of the four quadrants identified in Figure~\ref{Mix2}.  
Those pairs denoted as solid points are still located
close to a commensurability (within 1.50--1.55 or 2.00--2.05) at the end of the simulation,
 while those represented as open circles are well separated. 
The dotted lines indicate equal mass ratios for the inner pair (vertical) and outer pair (horizontal).
text. \label{mmp2}}
\end{figure}

Figure~\ref{mmp2} shows how the mass ratios influence the outcome of these models.
It shows the distribution of mass ratios for model planetary systems in
the four different quadrants.
Objects in quadrant A tend to have the outermost planet larger than the middle planet, and
those that are found far from resonance also tend to have the middle planet larger than the inner one.
The planets in quadrant B tend to be found with middle planets larger than inner planets. Those
planets in quadrant C tend to exhibit an overall trend of increasing mass outwards, while
those in quadrant D tend to have inner pairs with similar mass ratios. 

\begin{figure}
\centering
\includegraphics[width=1.0\linewidth]{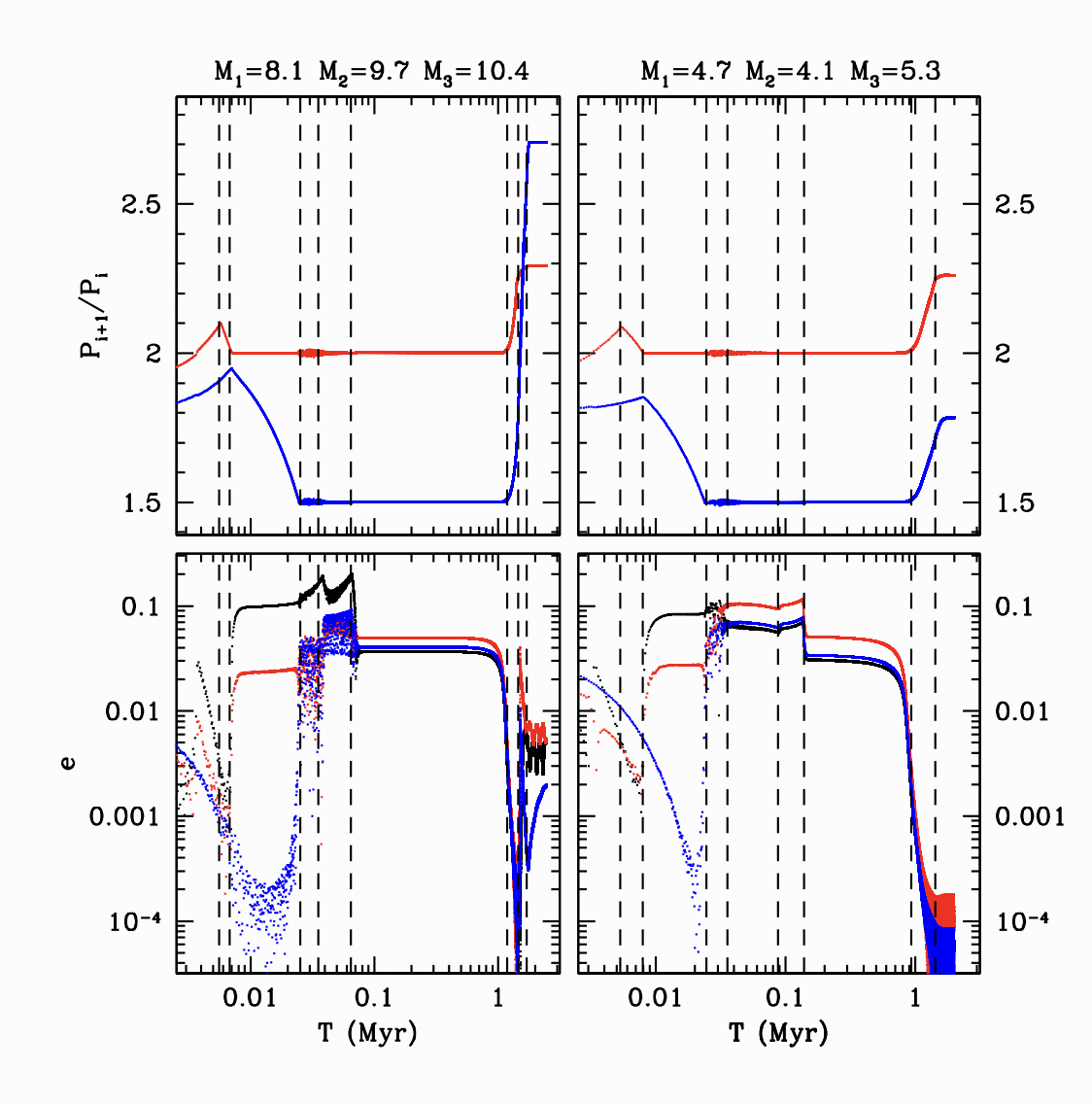}
\caption{  On the left hand side, $M_1=8.1 M_{\oplus}$, $M_2=9.7 M_{\oplus}$ and $M_3=10.4 M_{\oplus}$.  The colors identify
the planets in the same way as in Figure~\ref{Trip1}. We see
that this system starts in the same configuration as the system in Figure~\ref{Trip1}, but the outer planet diverges
so far that it crosses the 2:1 resonance, leaving both pairs wide of the 2:1 commensurability  and so located in quadrant C of Figure~\ref{Mix2}. In the lower left, we see that
the divergence across the 2:1 resonance results in an eccentricity excitation. On the right hand side, we see the evolution of a
three planet system  with  $M_1=4.7 M_{\oplus}$, $M_2=4.1 M_{\oplus}$ and $M_3=5.3 M_{\oplus}$. In this case the inner pair traps into the 2:1 and the outer pair into the 3:2 resonance. Their eventual divergence leaves the system in quadrant D.
 \label{Trip34}}
\end{figure}

 The examples shown in Figure~\ref{Trip1} and Figure~\ref{Trip2} demonstrate evolutionary pathways
that lead to quadrants A and B.
  Figure~\ref{Trip34} shows examples of
systems that end up in the other two quadrants, this time drawn from the Kepler analogs. The left hand panel shows a  relatively
massive system, with planets of mass 8.1, 9.7 and 10.5 $M_{\oplus}$, a realisation of
the masses in the system Kepler-60. It shows that the rapid inward migration causes the
inner pair to diverge to the point that they are wider than the 2:1 resonance. When the inner planet
halts  this trend reverses and the inner pair lock into the 2:1 resonance. This also slows
the inward migration of the middle planet so that the outer pair start to converge as well. However, this
happens just before they diverge beyond 2:1, so that the convergence ultimately leads to their 
trapping in the 3:2 resonance. This transition leads to  a period of dynamical rearrangement, as
the inner pair loses its resonant lock, before the outer pair actually locks into the 3:2 resonance and
only appears to settle down again once the outer planet reaches the torque reversal.

After this, the evolution is more stable, with a 2:1+3:2 chain evolving outwards until  eventually the system  diverges.
The larger planetary masses means that the system takes longer to decouple, and the outer pair passes through
the 2:1 resonance. The system doesn't trap because the evolution is divergent, but the resonant
crossing does produce an eccentricity excitation (observed in the lower left panel) that doesn't entirely die away before the system completely freezes out.
The end result is that neither pair remains in resonance at the end.


The right hand side of 
Figure~\ref{Trip34} shows the evolution of a planetary system modelled after Kepler-295, with planetary masses of
4.7, 4.1 and 5.3 $M_{\oplus}$ respectively, representative of  a system that ends up in quadrant~D. In this case, the evolution is quite similar to that seen in the left hand panel,
with the inner pair captured into 2:1 after an initial divergence, and the outer pair captured into the 3:2, with a similar
brief period of dynamical reshuffling. However, the outer planet is less massive in this case, and the
system does not evolve far enough to pass through the 2:1 commensurability.

\subsubsection{Initial conditions}
\label{KepWide}

Up to this point, we have assumed that the planets either formed close in or arrived from further out at an early stage, so
that their final states are determined by the outward migration at late times. However, if the planets spiral in at later times,
they may never reach the same compact configurations as we have simulated above. 

Thus, to test this assumption, we have evolved the same 107 Kepler analog systems forward, but now starting them at semi-major
axes of 0.16~AU, 0.29~AU and 0.54~AU.  This does indeed lead to wider final configurations, as many systems get trapped in
the 2:1 resonance during their inspiral, so that the detachment from resonance starts from a wider configuration.
Furthermore, systems with low mass outer planets may never experience resonance locking because the inner planets
migrate inwards faster and the outer planet cannot catch up before the disk evaporates. This can leave an inner pair
near the resonance accompanied by a very wide outer pair -- a system that will most often be observed simply as a double.

\begin{figure}
\centering
\includegraphics[width=1.0\linewidth]{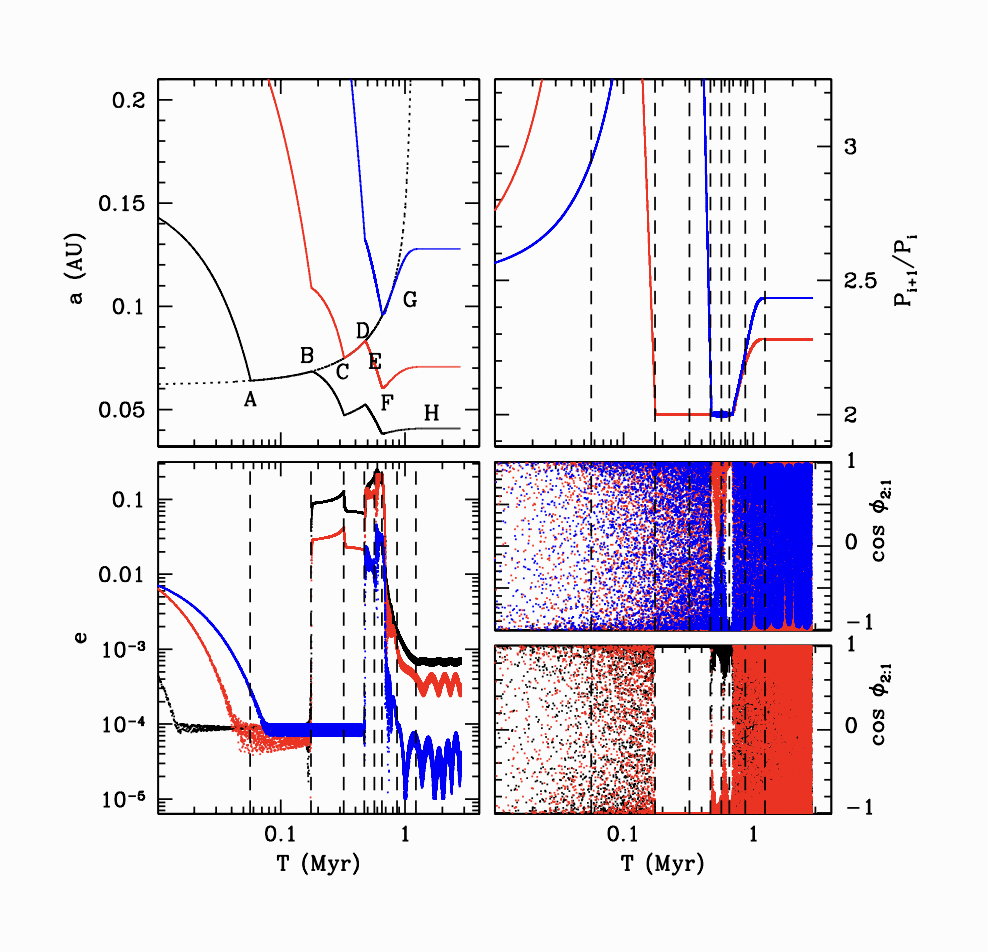}
\caption{ This three planet system is for $M_1=4.7 M_{\oplus}$, $M_2=4.1 M_{\oplus}$ and $M_3=5.3 M_{\oplus}$. 
The resonant angles at the lower right are for the 2:1 resonance for the inner pair (upper panel) and outer pair
(lower panel).  The system gets trapped into a 2:1+2:1 resonant chain that eventually diverges to a final
position in Quadrant C. \label{Trip5}}
\end{figure}

Figure~\ref{Trip5} illustrates the effect of late time inspiral by showing the evolution of a system with the same
mass configuration as in  the right hand side of  Figure~\ref{Trip34}, with the new, more distant, initial conditions. The inner planet
reaches the  torque reversal relatively early (at A), and the middle planet catches up to it at B. The systems started wider
in this case and so  both pairs capture into the 2:1 resonance at this point and the torques on the middle planet drive the
pair inwards until the middle planet reaches the torque reversal at C. Eventually, the outer planet gets close enough
to encounter the 2:1 resonance with the middle planet at D.

This causes some dynamical shake-up in the inner resonance but there is a period of time where both pairs
librate in a 2:1 resonance, albeit with non-zero amplitudes and changes in configuration (E). Eventually the outer planet reaches the
torque reversal (at F). This occurs late enough that the pairs immediately start diverging. Although both pairs pass through
period ratio 2.25 (at G), they are already out of resonance at this stage, and end up freezing out at H. At this point
the system is well located in quadrant C.

\section{Discussion}
\label{Disc}

The results of \S~\ref{Tres} demonstrate that our model results in the disruption of the majority
of the resonant chains that form initially, in keeping with the qualitative features of the observed Kepler sample.
To make a quantitative comparison, we must now compare the results of our simulations to the observations.

The Kepler satellite detects planets via their transits, so we must calculate the geometric probability that each
of the planets in our simulations are actually observed. This scales as the inverse of the semi-major axis. The relative frequency of transit of
members of multiple planet systems depends on the inclination dispersion within the system. To assess
these probabilities, we observe each system from $10^5$ random orientations on the sky. For each
realisation, the mutual inclinations are drawn from a Rayleigh distribution with an
inclination dispersion of $3^{\circ}$, the upper limit consistent with the Kepler statistics \citep{FM12}. 
The relative lines of nodes are assumed to be randomly distributed. With these results, we determine
the relative frequency of observing all three planets in each system, and the frequency of observing the
three different permutations of planetary pairs.

\begin{figure}
\centering
\includegraphics[width=1.0\linewidth]{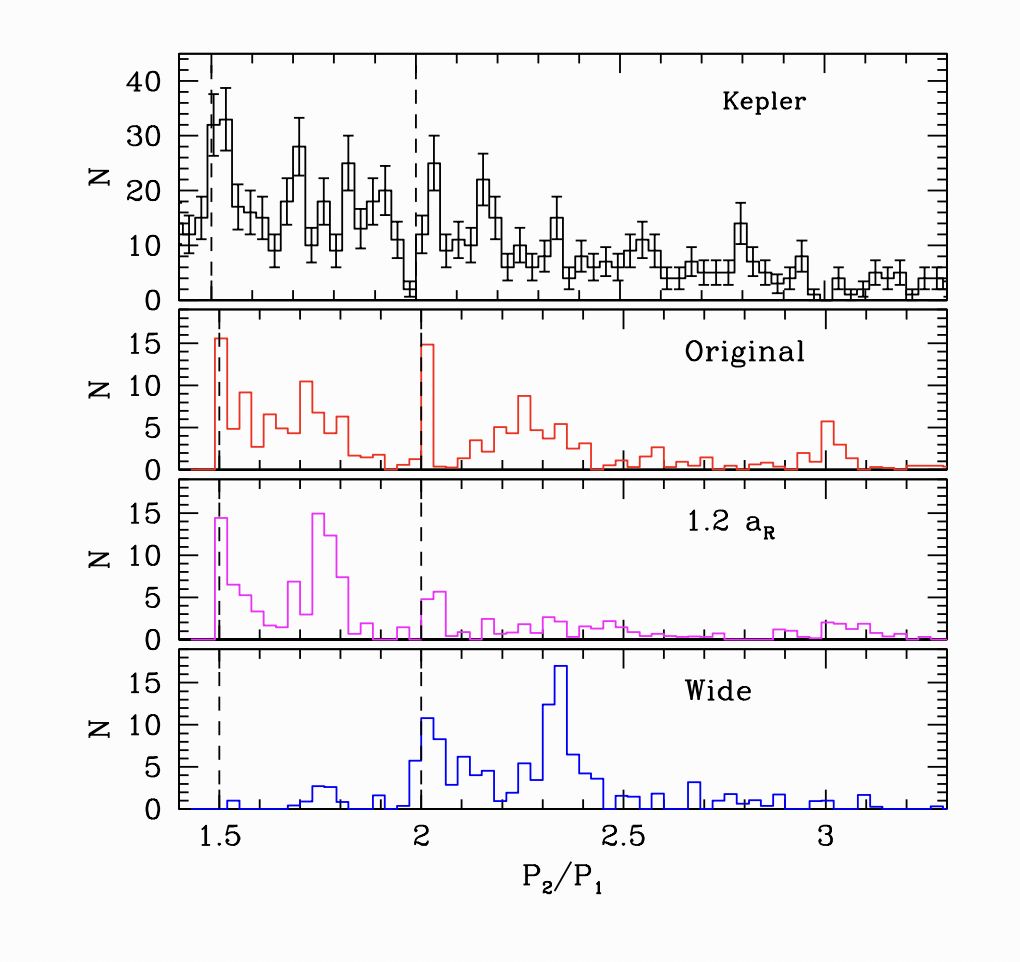}
\caption{ The upper panel shows the observed distribution of nearest neighbour period ratios in the
sample of Kepler systems, based on the data from the NASA exoplanet archive, with Poisson
error bars.
 The three panels below that show the outputs of our
simulations. The second panel (red histogram) shows the final distributions of the simulations described in \S~\ref{KepAn},
 which assumes that all stars the
default value of the transition radius $a_R$. The third panel (magenta histogram) shows the results of similar initial conditions, but 
 using 
 $a_R=1.2$, so that the migration halts at longer orbital periods.
The bottom panel (blue histogram)
shows the results from the simulations that began with the more distant, and wider spaced, initial conditions discussed in \S~\ref{KepWide}.
This means that more systems get trapped into the 2:1 resonance instead of the 3:2 resonances, and the
later divergence leads to wider systems. These simulations assume the standard $a_R$. \label{Prat}}
\end{figure}

Figure~\ref{Prat} shows the resulting  probability of observing different period ratios between neighbouring planets.
In the upper panel, we show the observed distribution of Kepler period ratios, using   those planets identified in the 
NASA Exoplanet Archive \citep{Nexsci} as confirmed, and discovered by Kepler.   The resulting distribution resembles
that of the standard  compilation, from \cite{Fab14}, but with improved statistics.
The two first
order resonances (q+1)/q, with q=1 and q=2, are shown as dashed lines, illustrating the well-known result that most
nearest neighbour pairs in the Kepler sample are not near a commensurability.

The other panels in Figure~\ref{Prat} show the output from our simulations. The second panel from the top, labelled as  `Original',
represents the results from our default simulations (the ones shown as red points in Figure~\ref{Mix2}). These
have been assigned weights according to their geometric probability of observation, as discussed above. The third
panel from the top represents the evolution of the same set of initial conditions, but with  $a_R$ scaled up by a factor of 1.2 relative
to the default
value given in equation~(\ref{TTrans}). This represents the case of a stronger stellar magnetic fields. The resulting distribution
is pretty similar, although the resonant peaks are less pronounced -- i.e. a larger fraction of systems moved away from
commensurabilities.

Finally, the lower panel shows the results of the simulations discussed in \S~\ref{KepWide}.
These start with  more widely separated initial conditions (and an $a_R$ corresponding
to that of the second panel, i.e. all systems have the default $a_R$). We see that many more of these systems get captured in the
2:1 resonance and we get a distribution that skews to wider separations than seen in the observations.

The most important feature of this figure is that the simulations produce a similar fraction of systems
wide of the commensurabilities as seen in the upper panel. The compact configurations produce a 
lot of systems between the q=1 and q=2 commensurabilities, while the more distant initial conditions
primarily produce systems wide of the q=1 commensurability.

\begin{figure}
\centering
\includegraphics[width=1.0\linewidth]{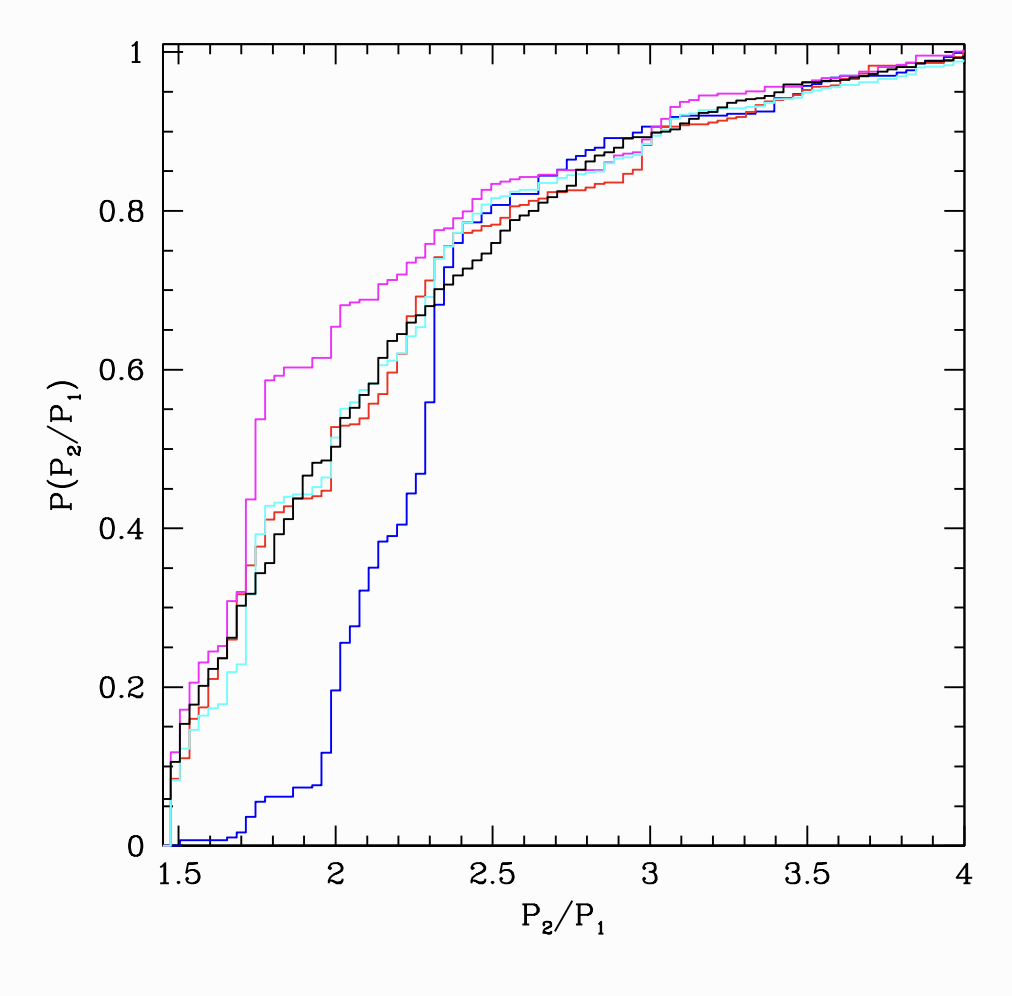}
\caption{ The  black histogram shows the cumulative distribution of period ratios for the
sample of current confirmed Kepler planetary systems -- the summed version of the solid
histogram in the upper panel of Figure~\ref{Prat}. The red histogram shows the cumulative
distribution from our default models -- those shown in the second panel of Figure~\ref{Prat}.
The magenta histogram shows the models from the third panel of Figure~\ref{Prat}, demonstrating
that the simulations with larger $a_R$ produce too many systems in compact configurations.
The blue histogram represents the lowest panel in Figure~\ref{Prat} -- those derived from the
simulations that start with wide initial conditions. The cyan histogram represents a sample
composed of 70\% from the magenta histogram and 30\% from the blue sample, indicating
that this too can produce a reasonably accurate representation of the observations.
  \label{Pcum0}}
\end{figure}

A more quantitative comparison of the non-resonant fractions  is shown in Figure~\ref{Pcum0},
which compares the cumulative distribution of period ratios between theory and observations,
 using the same colors as in Figure~\ref{Prat}.

 Thus, the black histogram represents the observed distribution drawn from the current Kepler sample (the
solid histogram in the upper panel Figure~\ref{Prat}). The red histogram represents the original model
sample from the second panel, and which reproduces the observed distribution quite accurately. The magenta histogram shows the distribution
from the third panel. This one shows too many at lower period ratios and too few at higher period ratios. 
The blue histogram shows the sample from the lowest panel -- those drawn from wide initial conditions.
In this case the model clearly produces too few close pairs and even falls below the observed
curve at larger periods, indicating a substantial pile-up of systems between period ratios of 2.2 and 3.
Overall, we find that the default model does an excellent job of providing a distribution of period
ratios broadly consistent with those observed. A Kolmogorov-Smirnov test indicates that the two
distributions have a 70\% chance of being drawn from the same underlying distribution.
The observations can also be reproduced with combinations
of the different samples. The cyan histogram shows a model in which 70\% of the systems come
from the simulations with larger $a_R$ and compact initial conditions, along with 30\% of the systems
that began with wider initial conditions.

It is instructive to compare these results with those of prior related studies. The magnetospheric
rebound model of \cite{LOL17} describes a model of divergent migration driven by a one-sided
torque exerted at the disk inner edge truncation, which can drive the outer planet away from an
inner companion and break the resonance. Simulations of the Kepler population using this
model \citep{LO17} does indeed produce a population of non-resonant planetary pairs, but still
leaves too many planets in resonant lock. Our model allows for multiple planets to experience
outwardly directed migration, and the divergence arises when the relative rates of migration start
to differ. Comparison of our simulation results with those of \cite{LO17} shows that our model
produces more widely spaced systems and a much better agreement with observation, as shown
in Figure~\ref{Pcum0}.

Another useful comparison set is the simulations  of \cite{Izi17} and \cite{IBR21}, the most comprehensive effort to date to compare the
migration model with observations in a quantitative fashion. \cite{Izi17} describe a model for the migration of low
mass planets in both static and turbulent disks, which decay away on Myr timescales, leaving planets trapped in
a chain of mean motion resonances. They furthermore run the simulations forward another 100~Myr after the dispersal
of the gas disk, and track the evolution of systems that undergo dynamical instabilities which can break the resonant
chains. The encouraging part of the results of \cite{Izi17} is that the resulting instabilities do indeed produce many
non-resonant systems. However, the fraction of resonant systems that remain is still far too high ($\sim 50\%$) and
the dynamical instabilities lead to planetary collisions which result in systems too massive compared to the observed systems.
The simulations of \cite{IBR21} takes these models further by including a prescription for gas-assisted pebble accretion
during the migration phase. Variation in the pebble flux and the formation time of planetesimals result in various levels
of dynamical instability. The authors find that it is possible to match the neighbour period ratio distribution of the Kepler planets
but, once again, only if one treats the fraction of unstable systems as a free parameter. For some parameter choices, most
of the simulated systems do go unstable, but these also, once again, produce systems that are too massive.

Our results show that both of these problems can be alleviated with a more physically motivated inner boundary condition.
We note that our approximate match to the observations in Figure~\ref{Pcum0} uses the full set of simulated systems, so
that our remaining resonant fraction is consistent with observations. Furthermore, these results are achieved using  planetary masses estimated directly from the observations, and so demonstrate that the period ratios and masses can be reproduced in a 
mutually consistent manner.  This approach does leave open the question of how and where these planets assembled, but
it demonstrates that their final orbital configurations are largely determined by the manner in which the disk evolves
and dissipates.

\begin{figure}
\centering
\includegraphics[width=1.0\linewidth]{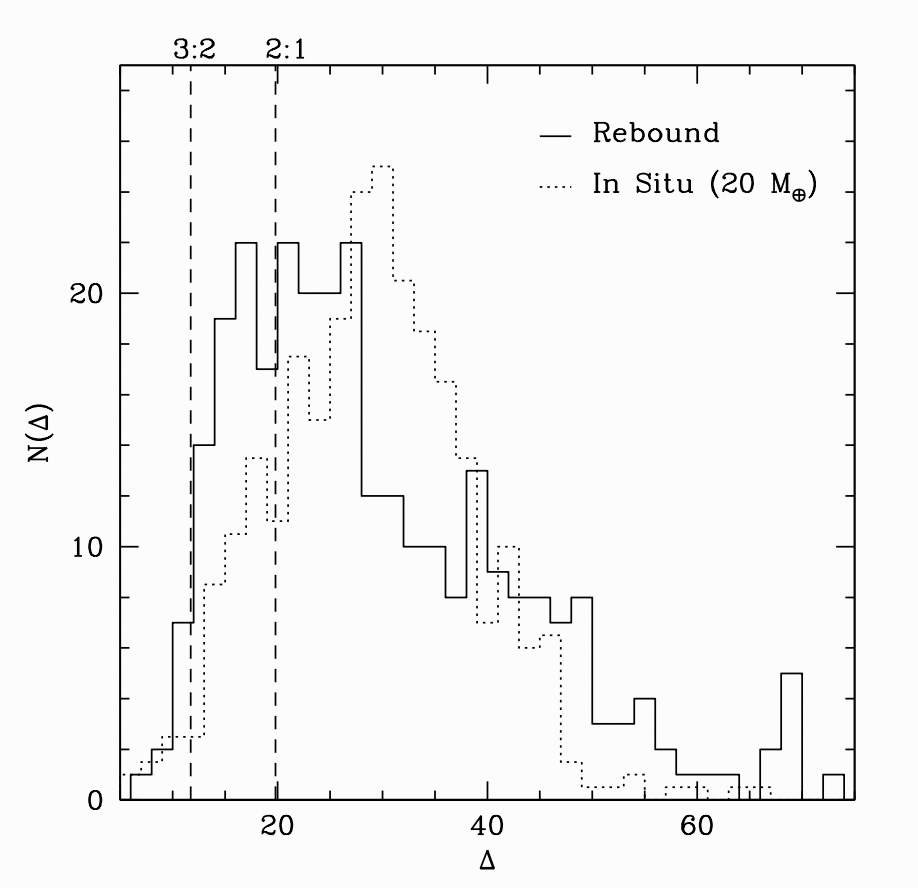}
\caption{ The solid histogram shows the spacings of the planets in our default set of 
Kepler analogue triples. The dotted histogram shows the mutual Hill spacings of planets emerging from
the in situ assembly simulations of \cite{HM13} (for a total solids inventory of 20 M$_{\oplus}$). The two
horizontal dotted lines indicate the spacings for a pair of 6$M_{\oplus}$ planets in either the 3:2 or 2:1 resonance.  
 \label{Hill}}
\end{figure}

The high multiplicity of these compact planetary systems has spurred many discussions regarding their
dynamical stability.
The question of whether the planetary systems are on the edge of dynamical instability  is often
expressed in terms of the planetary spacing as measured
in terms of mutual Hill radii.
It has been claimed that the Kepler multiplanet systems are close to being `maximally packed' \citep{LSM11,FM13,PW15}, in the sense that they
cannot be packed significantly closer without being dynamically unstable. Figure~\ref{Hill} shows the final spacings of 
our model three planet systems in terms of mutual radii. We see the final spacings of these planetary triples peaks
between 10 and 20 Hill radii, which is similar to that estimated for Kepler systems. 

 For comparison, the dotted histogram shows the outcomes from the in situ assembly simulations
of \cite{HM13} for the case of a total solid inventory of $20 M_{\oplus}$.  Both distributions show a broad spread
around a single peak, displaced by a factor of a few from the nominal stability limit  (\cite{PW15} and references therein).
The magnetospheric rebound model actually
produces systems that are even more compact than the in situ models, despite the fact that the latter are partially
sculpted by considerations of dynamical stability, while the spacing of the former is regulated by the freeze-out process of the gaseous disk.  As a consequence, it is misleading to infer that a closely packed planetary system is automatically the result of sculpting by late-stage dynamical instabilities.

\begin{figure}
\centering
\includegraphics[width=1.0\linewidth]{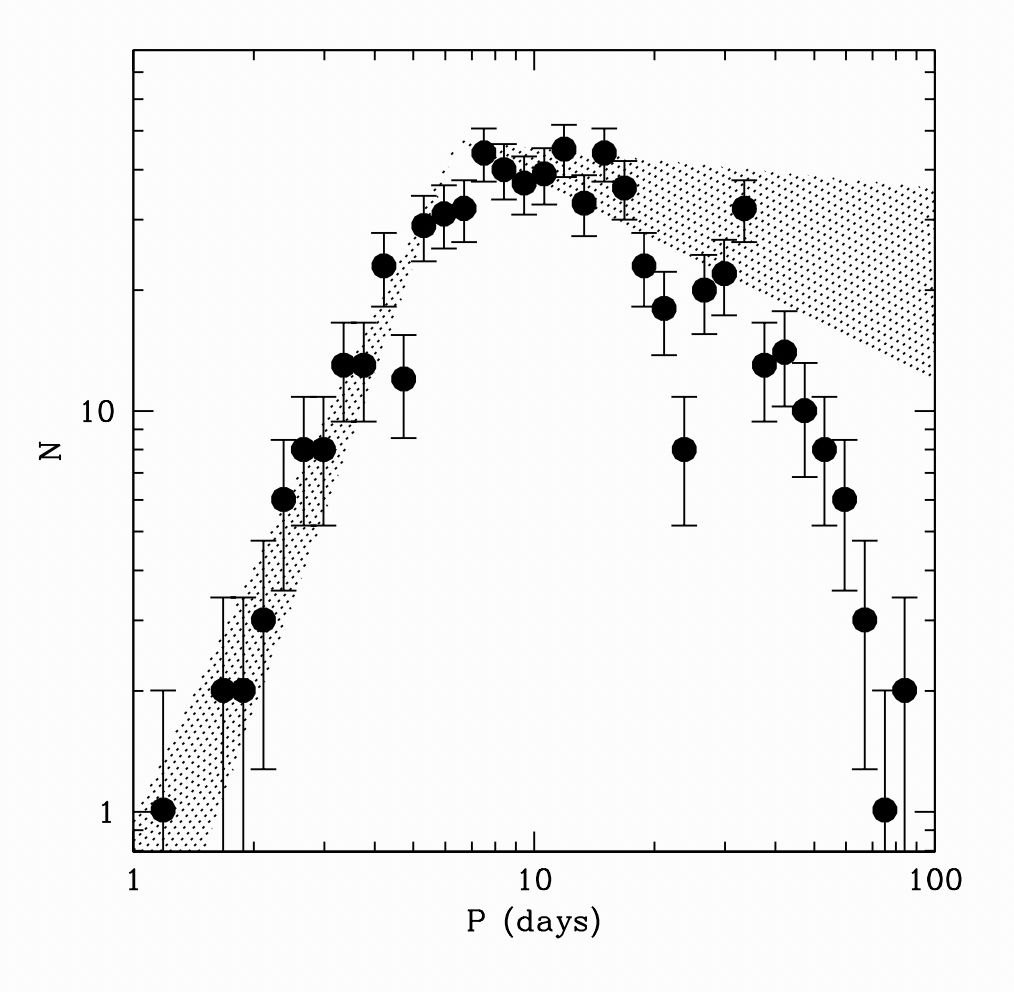}
\caption{ The solid points indicate the number of planets in our simulations that end their evolution at a 
given orbital period. The error bars are Poisson counting errors. The shade regions indicate the observationally
measured planetary occurrence rates for super-Earth planets from \cite{Peti18}. The shaded regions indicate the
range of slopes that lie within the observational uncertainties and the shaded regions have been scaled to match
at the peak. \label{Poccur}} 
\end{figure}

The time at which a given planet freezes out of the migration will also determine the final orbital period of the planet. This
will also be affected by the value of $a_R$ and hence the stellar magnetic field. To test whether our model values are
reasonable, we simulated the same 107 Kepler systems with the default value of $a_R$, a version with $a_R$ scaled up by
a factor 1.2 (discussed above already) and also one with a scaling factor of 0.8.
Figure~\ref{Poccur} shows the resulting distribution of final planetary locations for all planets summed over these three
simulations. In this case we do not adjust the counts for the geometric probabilities because the observations to which we
compare have already accounted for selection effects and purport to represent the true underlying distribution.
The shaded regions are these corresponding observed trends in planet occurrence for Super-Earths,
as measured by \cite{Peti18}. The simulations provide an excellent match to the observed rise from short periods to a peak in the range of $\sim$ 6--8 days,
followed by a drop-off towards larger periods. Immediately exterior to the peak, the simulations also  match the slope measured
by \cite{Peti18} but begin to drop below this at orbital periods $>$20~days. This is a reflection of our restriction to
triple systems. Systems with additional planets are expected to further fill in the distribution at longer periods.

It is particularly noteworthy that the slope of the rise is matched well by these simulations. Previous attempts to 
explain this power law behaviour include connecting it to a range in stellar rotation rates \citep{LC17}  via a
magnetically or centrifugally maintained
disk inner edge \citep{Koe91,KYH96, EHW05}. In the case of these calculations, it is determined by the mass distribution of
the underlying planetary systems, which will determine the degree to which a planet is forced interior to the torque
reversal and then when it freezes out. 

\section{Conclusions}
\label{Fin}

The results presented here demonstrate that a detailed treatment of the inner boundary of protoplanetary disks \citep{YHH23}
has a strong influence on the final configurations of planetary systems that migrate through this disk. In particular, these models
show the 
existence of an inner transition region of outwardly directed migration, which causes divergent migration of planets in resonant chains at late times. This results in a substantial fraction of  initially resonant chains diverging to the point that their period ratios no longer lie close to first order commensurabilities (although some may still be technically resonant because of their high precession rates). The resulting distribution of planetary neighbour period ratios is a good match to the observed distribution of planets detected by Kepler, and also provides a good match to the observed relative occurrence frequency of observed Super-Earth planets.
These results provide an explanation for the long-standing problem that the observed distribution of period ratios showed far fewer
resonant systems than expected in the models.

Our model is not as complete as some of the models that preceded it \citep{Izi17,IBR21}, in that we do not
present a full description of planetary evolution and assembly from planetary embryo formation up to final planetary architecture.
Rather, our model is designed to emphasize that some of the most important features of the observations (such as planetary orbital periods and neighbour period ratios) are likely to be the consequence of the structure of the inner protoplanetary disk, and how the planets evolve as the disk disperses. This is encouraging in the sense that it implies that the result is robust with respect to some of the
poorly understood physics involved in the assembly and early evolution of planetary systems.

On a less optimistic note, the fact that the period ratio distribution does not depend sensitively on the assembly process reduces the ways in which we can probe the formation conditions of planetary systems. It should be noted that our results match the observations if they begin with relatively compact configurations, so that examining the pathways to form  the initial configuration may still provide a fruitful avenue of investigation. We will address this issue in a future publication.

{\bf Data availability}: The data underlying this article will be shared on reasonable request to the corresponding author.

The authors thank the referees for prompt and comprehensive reports, which improved the exposition.
This research has made use of the NASA Exoplanet Archive, which is operated by the California Institute of Technology, under contract with the National Aeronautics and Space Administration under the Exoplanet Exploration Program. This research has made use of NASA's Astrophysics Data System Bibliographic Services.

\bibliographystyle{mnras}
\bibliography{refs}

\appendix

\section{A: Dynamical Equations}
\label{Eqns}

We wish to describe a system of three planets, subject to gaseous disk torques that cause migration in semi-major axis, and
damping of eccentricities. The planets also interact gravitationally with their nearest neighbours, and our equations include
the interactions with the first order resonances associated with the 2:1 and 3:2 commensurablities.

The planetary dynamics will be described in terms of the orbital frequencies, $n_1$, $n_2$ and $n_3$, normalised to a fidicial value $n_0$, so
that $x_i=n_i/n_0$. These correspond to semi-major axis $a_1<a_2<a_3$, so that $\alpha_{12}=a_1/a_2$ and
$\alpha_{23}=a_2/a_3$.
 Eccentricities are denoted by $e_i$ and the arguments of periastron are designed as $\bar{\omega}_i$. We will
incorporate effects of the first order resonances $(1+q)/q$ for $q=1$ and $q=2$, between nearest neighbours.  The  commensurate 
angles are $\psi_1=2 \lambda_2-\lambda_1$, $\psi_2=3 \lambda_2 - 2 \lambda_1$, $\psi_3=2 \lambda_3 - \lambda_2$
and $\psi_4 = 3 \lambda_3 - 2 \lambda_2$. We normalise the timescale to $t_0 = M_*/(3 n_0 m_2)$, so that $\dot{x}_i = 
dx_i/d\tau = t_0 dx_i/dt$.  The resonant angles are constructed by combining these commensurate angles with the relevant 
longitudes of periastron.

The resulting evolutionary equations \citep{MD2000,GS14,TP19} for the $x_i$ are
\begin{eqnarray}
\dot{x}_1 &=& - x_1^2 \alpha_{12} \left[ A_{12} \sin \psi_1
- B_{12} \cos \psi_1\right]  - 2 x_1^2 \alpha_{12}  \left[ C_{12}\sin \psi_2
- D_{12} \cos \psi_2\right]  + \frac{3}{2} \frac{x_1}{\tau_{a,1}} + 3 e_1^2 \frac{x_1}{\tau_{e,1}} \\
\dot{x}_2 &=& 2 x_2^2  \frac{m_1}{m_2} \left[ A_{12} \sin \psi_1
- B_{12} \cos \psi_1\right]  
+ 3 x_2^2 \frac{m_1}{m_2}  \left[ C_{12} \sin \psi_2
- D_{12} \cos \psi_2\right]+  \nonumber \\
&& - x_2^2 \alpha_{23} \frac{m_3}{m_2} \left[ A_{23} \sin \psi_3
- B_{23} \cos \psi_3\right]  
 - 2 x_2^2 \alpha_{23}  \frac{m_3}{m_2} \left[ C_{23} \sin \psi_4
- D_{23} \cos \psi_4\right]+ \frac{3}{2} \frac{x_1}{\tau_{a,1}} +
 3 e_2^2 \frac{x_2}{\tau_{e,2}} \\
  \dot{x}_3 &=& 2 x_3^2   \left[ A_{23} \sin \psi_3
- B_{23} \cos \psi_3\right]  + 3 x_3^2   \left[ C_{23} \sin \psi_4
- D_{23} \cos \psi_4\right]+  \frac{3}{2} \frac{x_3}{\tau_{a,3}} +
 3 e_3^2 \frac{x_3}{\tau_{e,3}}
\end{eqnarray}
and the coefficients $A$--$D$ are functions of the eccentricity,
\begin{eqnarray}
A_{ij} &=& f_1(\alpha_{ij}) k_i + f'_2(\alpha_{ij}) k_j \\
B_{ij} &=& f_1(\alpha_{ij}) h_i + f'_2(\alpha_{ij}) h_j \\
C_{ij} &=& f_3(\alpha_{ij}) k_i + f_4(\alpha_{ij}) k_j \\
D_{ij} &=& f_3(\alpha_{ij}) h_i + f_4(\alpha_{ij}) h_j 
\end{eqnarray}
and the quantities $f_i$ are the functions of the Laplace coefficients $b_{i}^{(j)} (\alpha)$, related to the expansion of the orbit-averaged
disturbing function in terms of the resonant arguments. These are given by
\begin{eqnarray}
f_1(\alpha) &=& - \frac{1}{2} \left( 4 + \alpha \frac{d}{d\alpha} \right) b_{1/2}^{(2)} (\alpha) \\
f_2(\alpha) & = &  \frac{1}{2} \left( 3 + \alpha \frac{d}{d\alpha} \right) b_{1/2}^{(1)} (\alpha) \\
f_3(\alpha) & = & - \frac{1}{2} \left( 6 + \alpha \frac{d}{d\alpha} \right) b_{1/2}^{(3)} (\alpha) \\
 f_4(\alpha) & = &   \frac{1}{2} \left( 5 + \alpha \frac{d}{d\alpha} \right) b_{1/2}^{(2)} (\alpha) \\
 f_2'(\alpha) & = & f_2 (\alpha) - 2 \alpha \\
 f_2''(\alpha) & = & f_2 (\alpha) - \frac{1}{2 \alpha^2} 
\end{eqnarray}
 The quantities $\tau_{a,i}$ represent the eccentricity damping timescales given by equation~(\ref{taua}), normalised to $t_0$. The
quantities $\tau_{e,i}$ represent the corresponding eccentricity damping timescales, taken to be $\tau_{e,i}=0.03 \left| \tau_{a,i} \right|$ (after fitting to the  calculations of \cite{YHH23}). Eccentricity damping for a given planet $i$ is switched off if $e_i<10^{-4}$. 
This eccentricity damping can often lead to low eccentricities which, in turn, leads to high precession rates. As a result,
some of the resonant angles can continue to librate even when the period ratios deviate from the exact
commensurability. Our equations represent the expansions in terms of the resonant arguments for the instantaneous osculating value
of $\alpha$, and so we describe the functions $f_j$ as functions of $\alpha$ and not simply a number at exact commensurability.

\begin{figure}
\centering
\includegraphics[width=1.0\linewidth]{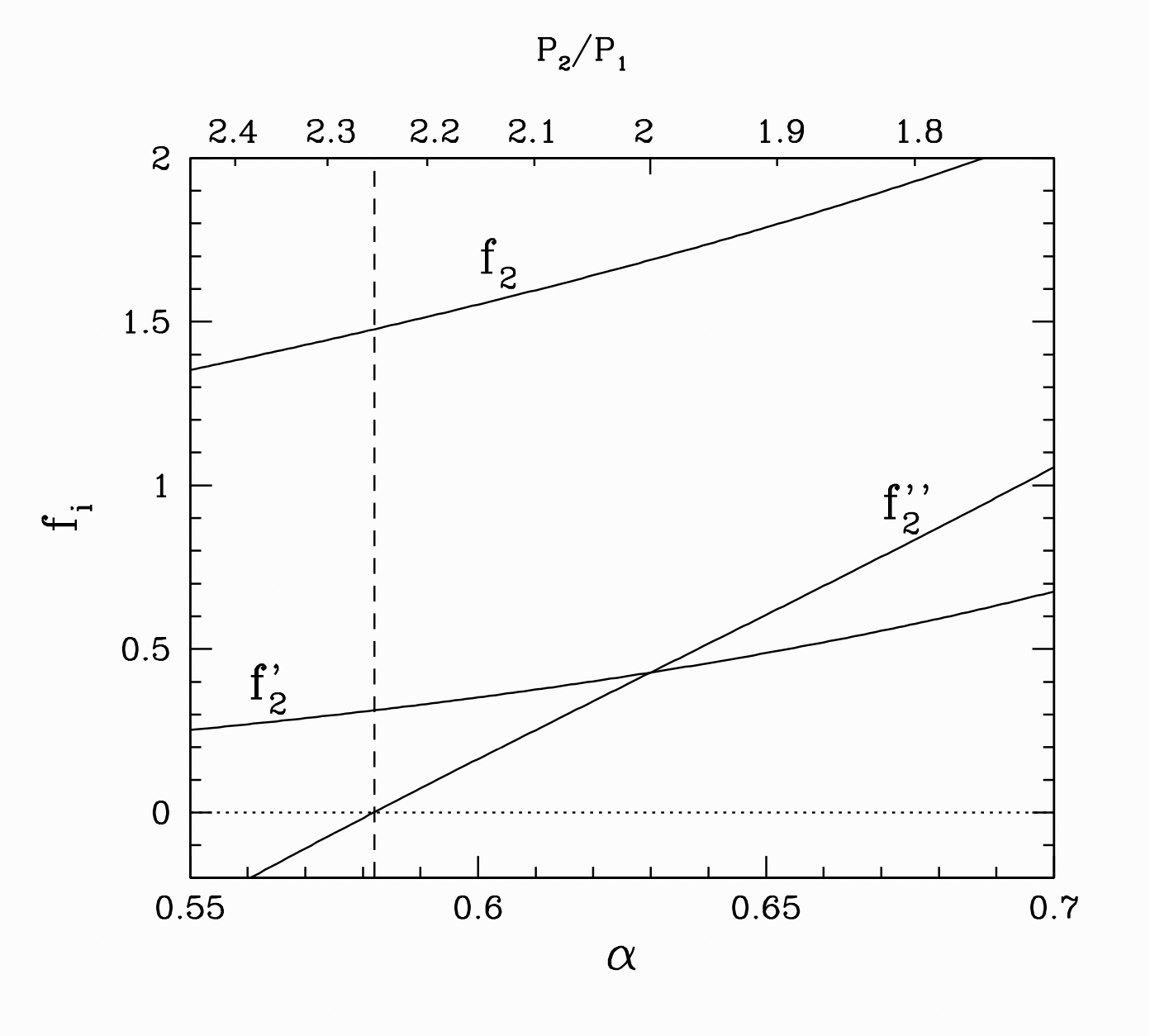}
\caption{ The solid curves indicate the functions $f_2(\alpha)$, $f_2'=f_2 - 2 \alpha$
and $f''_2 = f_2 - 1/(2 \alpha^2)$. The vertical dashed line indicates the value of $\alpha$ at
which $f''_2$ changes sign. The upper axis shows the corresponding period ratio, indicating that
the zero occurs at $P_2/P_1=2.252$. \label{f2}}
\end{figure}

 Another consequence of the low eccentricities is that it is prudent to replace $e_i$ and $\bar{\omega}_i$ with
$h_i$ and $k_i$, defined as
 $h_i = e_i \sin \bar{\omega}_i$ and $k_i = e_i \cos \bar{\omega}_i$. The resulting evolutionary equations are
\begin{eqnarray}
\dot{h}_1 & = & - \frac{h_1}{\tau_{e,1}} + \frac{\alpha_{12}}{3} f_1(\alpha_{12}) x_1 \cos \psi_1+\frac{\alpha_{12}}{3} f_3(\alpha_{12}) x_1 \cos \psi_2 \\
\dot{k}_1 & = & - \frac{k_1}{\tau_{e,1}} - \frac{\alpha_{12}}{3} f_1(\alpha_{12}) x_1 \sin \psi_1 + \frac{\alpha_{12}}{3} f_3(\alpha_{12}) x_1 \sin \psi_2 \\ 
\dot{h}_2 & = & - \frac{h_2}{\tau_{e,2}} + \frac{1}{3} \frac{m_1}{m_2}   f''_2(\alpha_{12}) x_2 \cos \psi_1 + \frac{1}{3} \frac{m_1}{m_2} f_4(\alpha_{12}) x_2 \cos\psi_2 \nonumber \\
&& + \frac{\alpha_{23}}{3} \frac{m_3}{m_2} f_1(\alpha_{23}) \cos \psi_3 + \frac{\alpha_{23}}{3}\frac{m_3}{m_2}  f_3(\alpha_{23}) x_2 \cos \psi_4 \\
\dot{k}_2 & = & - \frac{k_2}{\tau_{e,2}} - \frac{1}{3} \frac{m_1}{m_2} f''_2(\alpha_{12}) x_2 \sin \psi_1 - \frac{1}{3} \frac{m_1}{m_2} f_4(\alpha_{12}) x_2 \sin \psi_2  \nonumber\\
&& - \frac{\alpha_{23}}{3} \frac{m_3}{m_2}  f_1(\alpha_{23}) x_2 \sin \psi_3 + \frac{\alpha_{23}}{3} \frac{m_3}{m_2} f_3(\alpha_{23}) x_2 \sin \psi_4 \\
\dot{h}_3 & = & - \frac{h_3}{\tau_{e,3}} + \frac{1}{3}   f''_2 (\alpha_{23}) x_3 \cos \psi_3 + \frac{1}{3}  f_4 (\alpha_{23}) x_2 \cos \psi_4 \\
\dot{k}_3 & = & - \frac{k_3}{\tau_{e,3}} - \frac{1}{3}  f''_2(\alpha_{23}) x_3 \sin \psi_3 - \frac{1}{3}  f_4(\alpha_{23}) x_3 \sin \psi_4 
\end{eqnarray}

Finally, the above equations depend on the commensurabilities $\psi_1 = 2 \lambda_2 - \lambda_1$, $\psi_2 = 3 \lambda_2 - 2 \lambda_1$,
$\psi_3 = 2 \lambda_3 - \lambda_2$ and $\psi_4 = 3 \lambda_3 - 2 \lambda_2$. The evolution of these variables is given by
\begin{eqnarray}
\dot{\psi}_1 & = & \frac{1}{\Delta} \left( 2 x_2 - x_1\right) \\
\dot{\psi}_2 & = & \frac{1}{\Delta} \left( 3 x_2 - 2 x_1 \right) \\
\dot{\psi}_3 & = & \frac{1}{\Delta} \left( 2 x_3 - x_2 \right) \\
\dot{\psi}_4 & = & \frac{1}{\Delta} \left( 3 x_3 - 2 x_2 \right) 
\end{eqnarray}
where $\Delta = 3 m_2/M_*$.  The true resonant angles $\phi$ are determined by adding the relevant longitudes of periastron to the corresponding commensurabilities.

One technical detail that results from our use of the full functional form of $f_i(\alpha)$ is that the function $f''_2$ 
changes sign if $\alpha<0.582$, which corresponds to a period ratio $>2.252$. This is shown in Figure~\ref{f2} and
is a consequence of the indirect contributions to the disturbing function.
The sign of $f''_2$ determines the stable libration configuration of the 2:1 external resonance and so we see, in
some of our results, a flip of the resonant libration as the period ratio increases. This only occurs if the resonance
is well separated from the instability, which requires a fast libration and thus a small eccentricity.  As a result, it is not clear this
has an observable consequence but we include it for completeness.

\end{document}